\documentclass[11pt,a4paper]{article} 
\usepackage{jheppub}
\usepackage{enumerate}
\usepackage{wasysym} 
\usepackage{mathrsfs} 
\usepackage{graphicx} 
\usepackage{amsfonts} 
\usepackage{amsbsy} 
\usepackage{amscd}
\usepackage{slashed} 
\usepackage{multirow} 
\usepackage{color}
\definecolor{myred}{rgb}{0.6,0,0} 
\definecolor{myblue}{rgb}{0,0.2,0.4}
\definecolor{mygreen}{rgb}{0,0.9,0.1}
\definecolor{hc}{rgb}{.9,0.1,0.7}
\definecolor{hcout}{rgb}{.9,0.7,0.9}
\definecolor{Orange}{rgb}{1.,0.65,0.}

\makeatletter

\newcommand{\fmslash}[2][0mu]{%
  \mathchoice
    {\fmsl@sh\displaystyle{#1}{#2}}%
    {\fmsl@sh\textstyle{#1}{#2}}%
    {\fmsl@sh\scriptstyle{#1}{#2}}%
    {\fmsl@sh\scriptscriptstyle{#1}{#2}}}
\newcommand{\fmsl@sh}[3]{%
  \m@th\ooalign{$\hfil#1\mkern#2/\hfil$\crcr$#1#3$}}
\makeatother

\newcommand{\lsim}{{\;\raise0.3ex\hbox{$<$\kern-0.75em\raise-1.1ex\hbox{$\sim$}}\;}}
\newcommand{\gsim}{{\;\raise0.3ex\hbox{$>$\kern-0.75em\raise-1.1ex\hbox{$\sim$}}\;}}

\newcommand{\D}{\fmslash D}
\newcommand{\Del}{\fmslash \partial}

\usepackage{array}
\newcolumntype{C}[1]{>{\centering\arraybackslash$}p{#1}<{$}}

\usepackage{bm} 
\newcommand{\be}{\begin{equation}}
\newcommand{\ee}{\end{equation}}
\newcommand{\bes}{\begin{equation*}}
\newcommand{\ees}{\end{equation*}}
\newcommand{\bea}{\begin{eqnarray}}
\newcommand{\eea}{\end{eqnarray}}
\newcommand{\beas}{\begin{eqnarray*}}
\newcommand{\eeas}{\end{eqnarray*}}
\title{Models of decaying FIMP Dark Matter: potential links with the Neutrino Sector} 

\author[a]{Laura Covi,}
\author[b]{Avirup Ghosh,}
\author[c]{Tanmoy Mondal,} 
\author[d]{Biswarup Mukhopadhyaya} 
\affiliation[a]{Institute for Theoretical Physics, Georg-August University G\"{o}ttingen,
Friedrich-Hund-Platz 1, G\"{o}ttingen, D-37077 Germany}
\affiliation[b]{Regional Centre for Accelerator-based Particle Physics,
Harish-Chandra Research Institute, HBNI,
Chhatnag Road, Jhunsi, Allahabad - 211 019, India}
\affiliation[c]{School of Physics, Korea Institute for Advanced Study, Seoul 02455, South Korea} 
\affiliation[d]{Department of Physical Sciences, Indian Institute of Science Education and Research, Kolkata, Mohanpur - 741246, India} 
\emailAdd{laura.covi@theorie.physik.uni-goettingen.de}
\emailAdd{avirupghosh@hri.res.in}
\emailAdd{tanmoy@kias.re.kr}
\emailAdd{biswarup@iiserkol.ac.in}

\abstract{The absolute stability of a dark matter (DM) particle is not a
binding requirement. Here we suggest a few scenarios where the DM particle
is liable to decay via extremely feeble interactions.  This can happen via
inexplicably small Yukawa couplings in the simplest conjectures. 
After setting down such a model, we go beyond it, thus treading onto scenarios where 
the spontaneous breakdown
of some gauged $U(1)$ symmetry may lead to intermediate scales, and suitably
suppressed effective operators which allow the DM particle to decay slowly. The constraints
from particle physics as well as cosmology are taken into account in each case. The last and
more involved scenario, studied in detail, suggest a link between the model parameters that govern 
neutrino physics on  one side, and the dynamics of a quasi-stable DM particle on the other. }

\preprint{HRI-RECAPP-2020-006 \\$\textrm{}$\hfill KIAS-P20044 \\$\textrm{}$\hfill \today}


\begin{document}
\maketitle

\newpage

\section{Introduction} \label{sec:intro}
Dark matter (DM) is an undeniable component of the universe today, playing a fundamental role in 
structure formation and in explaining galactic rotation
curves and other astrophysical and cosmological observations~\cite{Ade:2015xua}.
Assuming a $\mathbb{Z}_2$ symmetry is a frequently adopted practice
in ensuring a stable particle in the elementary particle spectrum, 
which can account for dark matter (DM) in our universe. In
special cases like the minimal supersymmetric standard 
model (MSSM) lepton and baryon number 
conservation and stability of the proton may (though somewhat 
grudgingly) be taken as facts supported by experiments. In general,
however,  such broader theoretical motivation for  
$\mathbb{Z}_2$ symmetries are  difficult to find.
Furthermore, global symmetries are not likely to be 
respected by quantum gravity \cite{Banks:2010zn,Mambrini:2015sia,Harlow:2018jwu}.
Thus even a scenario that is $\mathbb{Z}_2$-symmetric at low energy may permit
very small violation of the discrete symmetry, when one takes its UV-completion
into account.

On the other hand, dark matter does not have to be absolutely stable.
Indeed it is possible that the dark matter candidate(s) has extremely slow 
decays, with a lifetime much longer than the age of the Universe, due
to correspondingly small couplings, whose smallness is protected
in spite of radiative corrections. Here we wish to illustrate
such a  scenario, devoid of any discrete symmetry and consisting of 
a very long-lived dark matter particle.
For these very small DM couplings, the most typical mechanisms for dark matter
production are the freeze-in \cite{Hall:2009bx, Bernal:2017kxu} and SuperWIMP mechanisms \cite{Covi:1999ty,Feng:2003xh}, which
rely on the generation of the DM particles from a mother particle in thermal equilibrium. In our model we will 
therefore also have additional charged states, which can be in equilibrium with the SM, produce DM and mediate 
interactions between the DM and the dark sector.

The approach to construct a model with decaying dark matter, 
followed in this work, consists of three levels with increasing complexity of the model as well as naturality. 
We will consider in all cases a spin-1/2 DM candidate which can decay
only via  very small Yukawa interactions or higher-dimensional operators. 

As the first case, we consider a model with minimal field content and renormalizable Yukawa couplings 
driving DM decay. It should be remembered here that the Yukawas in the standard
model (SM) vary over some five orders of magnitude, without any fundamental principle explaining them. 
Though this is somewhat dissatisfying, a redeeming feature is that these couplings are `$\textit{technically natural}$'~\cite{tHooft:1979rat},
since their radiative corrections, are always proportional to
the tree-level Yukawa couplings with additional coefficients $\le 1$. Emboldened by this, we construct a scenario 
with not only new Yukawas of even smaller magnitude than those in the SM, but also some gauge-invariant fermion 
masses, which are all shown to be  stable against 
radiative corrections for certain ranges of values of the parameters. This again makes the added terms {\it `technically natural'}. 

This `simple' scenario leads not only to a dark matter candidate consistent 
with relic density from $\textit{freeze-in}$, but also to an entire spectrum consistent with 
neutrino masses and mixing, FCNC, lepton universality, Higgs decay data etc. 
We introduce in the model three SM singlet Majorana fermions, 
the lightest of which serve as the dark matter candidate, bringing this model
within the class of decaying sterile neutrino DM models similar to the $\nu$SM~\cite{Asaka:2005pn, Shaposhnikov:2006xi,Shaposhnikov:2006nn}.

In addition a vector-like doublet $F$ has been considered, whose decay is responsible for the 
$\textit{freeze-in}$ production of the dark matter, as in~\cite{Arcadi:2013aba, Arcadi:2014tsa}.
The dark matter decays to three fermions via Yukawa interaction with the SM Higgs, 
the strength of which needs to be extremely small ($\le 10^{-20}$) in order to be consistent 
with DM decay observables. The smallness of this interaction strength, of a degree much more 
severe than what is seen in the SM, is inexplicable from the premises of the model, even
if radiatively stable.

To take care of the above issue, a slightly expanded scenario is proposed in the next step.  
We add a local $U(1)$ symmetry that is broken spontaneously with the help of a scalar 
$\phi$ at an intermediate scale around $10^{8}$ GeV. All the dark sector fields, $F, \psi$(DM) and $\phi$,
 are charged under this new gauge symmetry. 
One can write down dimension 5-and-6 operators, invariant under the SM gauge group as well 
as the new $U(1)$, suppressed by the Planck scale. Once the $U(1)$ symmetry is broken spontaneously,  
these higher-dimensional operators lead to mixing and highly suppressed interaction terms between 
the dark sector fermions and the SM leptons. These not only generate the tiny Yukawa couplings 
that causes the DM to decay  but also the interactions and the decay amplitudes for the $F$ to decay 
into DM with the level of smallness consistent with $\textit{freeze-in}$ production.

Finally, we upgrade the $U(1)$ gauge symmetry to $U(1)_{L_\mu-L_\tau}$ so as 
to establish a direct connection between DM and the leptonic sector. It is well-known
that $ U(1)_{L_\mu-L_\tau}$  can provide an explanation for the large neutrino mixing in the 
$ \mu-\tau $ sector \cite{Baek:2001kca,Lam:2001fb,Ma:2001md}.
Moreover quite a number of DM models have already been put forward in the context of
$U(1)_{L_\mu-L_\tau}$~\cite{Baek:2008nz,Baek:2015fea,Patra:2016shz,Altmannshofer:2016jzy,Biswas:2016yjr,Biswas:2016yan,Biswas:2017ait} 
but often in different contexts than in the present paper.
This scenario has all the advantages of the earlier model along with 
the DM-neutrino connection which provides it some additional merit. 
Thus we present the phenomenology of this model in greater 
details. Here the presence of the higher-dimensional terms in the
neutrino mass matrix allow us to obey the PLANCK bound on sum of the 
light neutrino masses  unlike the simple case with only renormalizable 
terms~\cite{Asai:2017ryy,Asai:2018ocx}. 
Moreover we have observable predictions for neutrinoless double-beta decay $0\nu\beta\beta$. 

The last-mentioned `gauged scenario'  may also be motivated from the angle of UV-completion.
On the one hand, such a symmetry is attractive from a neutrino physics point of view.
On the other, such a $U(1)$ may be the result of the breaking chain of a gauge group corresponding to a grand unified theory (GUT) 
at an intermediate scale~\cite{QUIROS1987461,PhysRevLett.60.1817}. Thus both a quasi-stable DM and the physics of lepton sector can be linked to a GUT scenario.

The paper is organized as follows.
In section~\ref{sec:simpmod} we discuss the $U(1)$-model  followed by 
a short discussion of the renormalizable model.  In 
section~\ref{sec:mutaumodel} we discuss the $L_\mu-L_\tau$ scenario in details.
We summarize and conclude in section~\ref{sec:conclusion}.

\section{Simplified Models}
\label{sec:simpmod}
\subsection{Model 1}
We first consider a model with three generations of RH fermions and one $SU(2)_L$ doublet vectorlike fermion in addition to the the SM particle content without imposing any additional symmetries. The newly added particles and their respective charges is shown in tab.~\ref{tab:qunum} where three generations of singlet fermions $N_{R,1},N_{R,2},N_{R,3}$ are collectively denoted as $N_{R,i}$. 
\begin{table}[t]
\begin{center}
\begin{tabular}{|c|c|c|c|}
\hline
Fields & $SU(3)_{c}$ & $SU(2)_{L}$ & $U(1)_{Y}$  \\
\hline
$N_{R,i}$ & 1 & 1 & 0  \\
$F=\left(\begin{matrix}
 F^{0} \\
 F^{-}
\end{matrix}\right)$  & 1  & 2 & -1/2  \\
$\overline{F}=\left(\begin{matrix}
 \overline{F}^{0} \\
 F^{+}
\end{matrix}\right)$  & 1  & $\overline{2} $ & 1/2  \\
\hline
\end{tabular}
\caption{The quantum numbers of the new fields in Model 1.}
\label{tab:qunum}
\end{center}
\end{table}

The renormalizable Lagrangian comprised of the newly added fields is,
\begin{eqnarray}
\mathcal{L}=&&\overline{N}_{R,i}i\D\,N_{R,i}+\overline{F}i\D\,F-M_{N,i}\overline{N^{c}_{R,i}}N_{R,i}-M_{F}\overline{F}F
-M_{l_i F}\left(\overline{L_{L,i}}F_{R}+\overline{F_{R}}L_{L,i}\right)\nonumber\\
&-&\left(Y_{N_i}\overline{L_{L}}H^{c}N_{R,i}+Y^{\dagger}_{N_i}\overline{N_{R,i}}H^{\dagger}L_{L}\right)- Y_{N_{i} F}\left(\overline{F_{L}}H^{c}N_{R,i}+\overline{N_{R,i}}H^{\dagger}F_{L}\right)\nonumber\\
&-&Y_{e_i F}\left(\overline{F_{L}}H\,l_{R,i}+\overline{l_{R,i}}H^{\dagger}F_{L}\right).
\label{eqn:model1}
\end{eqnarray}

\begin{figure}[t]
 \begin{center}
 \includegraphics[width=14cm,angle=0]{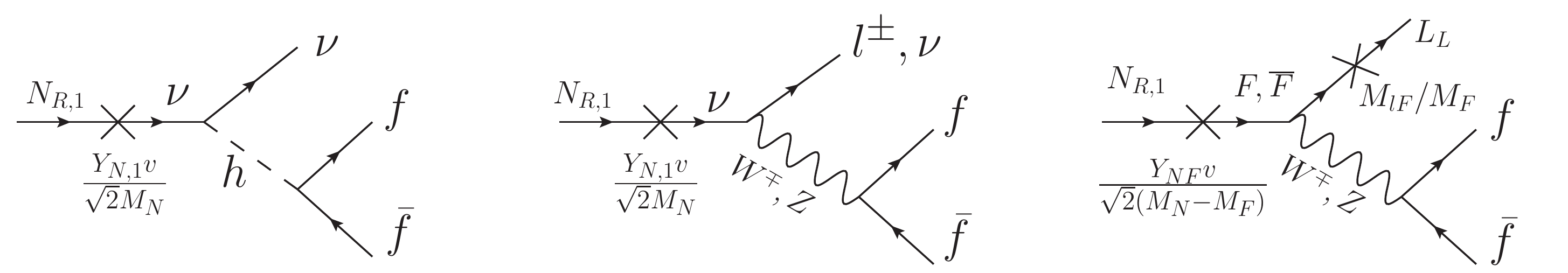}
 \caption{Feynman diagrams of DM decay in Model 1. The first two are the dominant processes to drive the decay of the DM $N_{R,1}$,
 the last decay channel goes via virtual vector-like fermion through the mixing $ M_{lF} $.}
 \label{fig:N1decay1}
 \end{center}
 \end{figure}

We have considered the lightest of the $N_{R,i}$, $N_{R,1}$ to be the DM candidate which mixes with the SM neutrinos and 
the new vectorial fermions and thus decays as shown in fig.~\ref{fig:N1decay1}. 
Also the decay channel via loops into neutrino and photon is present, but it is negligible for DM masses
above the 3 lepton decay threshold. 

The DM can be produced from the decay of $F, \overline F$ fermions via the $\textit{freeze-in}$ mechanism, 
as long as the relevant Yukawa coupling $ Y_{N_1 F} $ is in the range $\sim 10^{-11}-10^{-12} $, according to
\begin{eqnarray}
\Omega_{FI} h^2 &\sim & \frac{1.09 \times 10^{27} g_F}{g_*^{3/2}}  \frac{M_{N,1} \Gamma_{F\rightarrow \psi H}}{M_F^2}
=  0.1    \left( \frac{ g_* }{10^2} \right)^{-3/2} \left( \frac{ Y_{N_1 F}}{3.78\times 10^{-12}} \right)^2 \frac{ M_{N,1}}{M_F}, 
\label{eqn:FIMP-model1}
\end{eqnarray}
where $ g_F (= 4)$ counts the number of degrees of freedom in the $ F $ doublet, $ g_* $ is the number of relativistic degrees
of freedom in the thermal bath at the time of decay~\footnote{We are assuming here no entropy production between the
FIMP production and the present epoch.}. 

\begin{figure}[t]
 \begin{center}
 \includegraphics[width=7cm,angle=0]{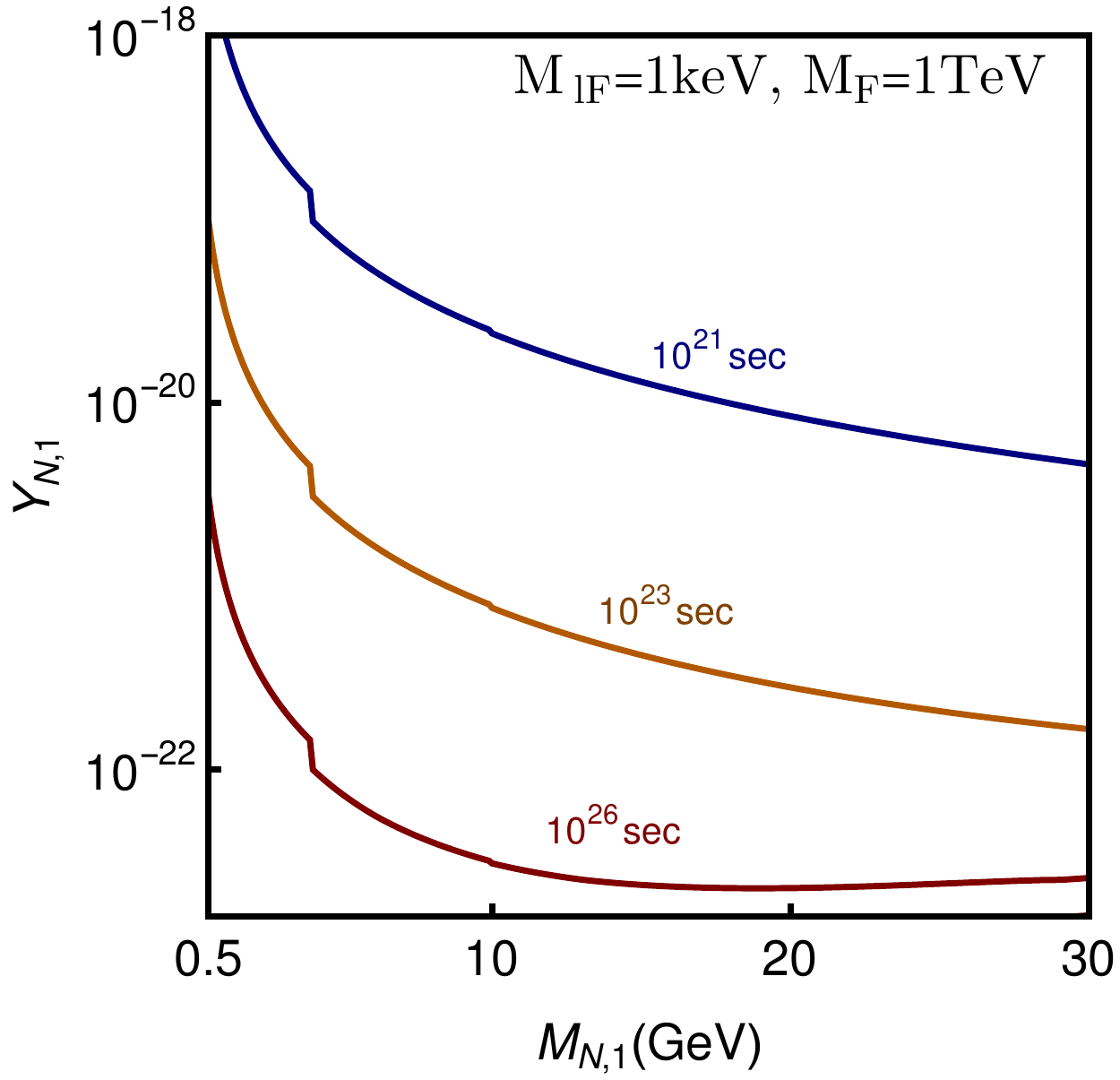}
 \caption{Contours of DM($N_{R,1}$) lifetime in the $M_{N,1}-Y_{N_1}$-plane.}
 \label{fig:N1decay2}
 \end{center}
 \end{figure}

Indirect detection constraints put a lower bound of $\mathcal{O}(10^{26}\,{\rm sec})$ on the lifetime of a decaying DM~\cite{Essig:2013goa}. 
In fig.~\ref{fig:N1decay2} we depict the lifetime contours of the DM as a function of DM mass and coupling. 
Evidently, in order to satisfy this constraint one needs very small Yukawa coupling $ Y_{N_1} $ as well as 
a very small mixing $M_{lF} \lesssim 1\,{\rm keV}$ to ensure a suppressed decay rate.

This value of $M_{lF}$ is much smaller than the mass of the heavy fermions and to check its stability we computed explicitly the one loop radiative 
contribution to $M_{lF}$. The corrections are as follows:

\begin{eqnarray}
\Delta\,M_{F^{0}\nu} \sim \left \{ 
   \begin{aligned}
   & g^{2}_{2} \times\,\mathcal{F}(p^2,m_{S,V},m_f) \times M_{lF} ~~\text{(F-V mediated loop)} \\
   &  g^{2}_{2} \times\,\mathcal{F}(p^2,m_{S,V},m_f) \times \frac{Y_{N}Y_{NF}\,v^{2}}{M_{N}-M_{F}} ~~\text{($N_i$-V mediated loop)} \\
   & Y_{N}\,Y_{NF}\,M_{N}\,\times\,\mathcal{F}(p^2,m_{S,V},m_f)~~\text{(H-$N_i$ loop)}.
   \end{aligned}
   \right.
\end{eqnarray}
and
\begin{eqnarray}
\Delta\,M_{F^{\mp}l^{\pm}} \sim \left \{ 
   \begin{aligned}
   & g^{2}_{2} \times \mathcal{F}(p^2,m_{S,V},m_f) \times M_{lF} ~~\text{(F-V loop)} \\
   & g^{2}_{2} \times \mathcal{F}(p^2,m_{S,V},m_f) \times \frac{Y_{N}Y_{NF}\,v^{2}}{M_{N}-M_{F}} ~~\text{(N$_i$-V loop)} \\
   &  Y_l M_{lF} \times \mathcal{F}(p^2,m_{S,V},m_f) ~~\text{(H-F loop).} 
   \end{aligned}
   \right.
\end{eqnarray}
Here  $\mathcal{F}(p^2,m_{S,V},m_f)=-\dfrac{1}{16\pi^2}\overset{1}{\underset{0}{\int}}dx \log\Delta(x)$ and $\Delta(x)=x m^2_{S,V}+ (1-x) m^2_{f} - x(1-x)p^2$ 
with $m_{S,V,f}$ being the mass of the scalar, gauge boson or fermion in the loop and $p$ denotes the incoming momentum. 
The parentheses in each case denote the particles running in the corresponding loops where $F = F^0,F^-$, $V=W^{\pm},Z,\gamma$ and $H$ denotes the 
SM higgs doublet.

So, if $Y_{NF}\,\leq\,10^{-10}$ then the corrections are too small to change substantially the mixing and affect the phenomenology. 
The coupling $Y_{N_1 F}$ has to be $\mathcal{O}(10^{-12})$ for freeze-in mechanism and it is natural to assume all $Y_{N_i F}$ couplings 
are of same order, producing negligible radiative corrections. 

The neutrino masses can be explained via type-I seesaw mechanism with appropriate values of $Y_{N,2,3}$ and $M_{N,2,3}$,
leaving one vanishing mass eigenstate.  For RH neutrino masses $M_{N,2,3}$ around $ 10-100 $ GeV also those Yukawas are small, 
below $ 10^{-7} $, but substantially larger than all the others.
So for the neutrino sector, the model is similar to the $\nu$SM model \cite{Asaka:2005pn, Shaposhnikov:2006xi,Shaposhnikov:2006nn}, 
and indeed it may be possible to produce here
also the baryon asymmetry of the Universe through the oscillations of the $ N_{2,3} $ states. On the other hand in this case
the production of the lightest heavy neutrino $ N_1 $ goes via another production mechanism than the Fuller-Shi mechanism~\cite{Shi:1998km}
and does not have to rely on the presence of a very large lepton asymmetry. 



Some positive and negative aspects of this simple-minded scenario are:
\begin{itemize}
\item It  is simple and minimalistic, and postulates no additional
charge, discrete or continuous, for the DM particle.

\item The scenario is technically natural. It is shown by explicit
calculation that not only the Yukawas but also the additional,
bare mass terms involving $F$ are stable against radiative corrections
for certain regions in the parameter space. 

\item However, justifying the ultra-small Yukawa interactions $ \sim 10^{-20} $, and
explaining why they are not zero to start with, is a potential difficulty. 
Also it appears that there are three different Yukawa coupling sizes, related to
the neutrino masses, the \textit{freeze-in} mechanism and the DM decay,
which have to be chosen ad-hoc.

\item $M_{lF}$ are `\textit{technically natural}' but being vectorlike bare mass terms one would 
naturally expect them to be in the same ballpark as $M_F$. But constraints on DM decay forces 
$M_{lF} \leq 1\,{\rm keV}\,\ll M_F$ which is difficult to explain.   
\end{itemize}

%

\subsection{Model 2}
In the second model we have the same fermions as in the previous one, but we have added a 
new $U(1)_{DM} $ gauge group and one charged scalar field ($\Phi$) which breaks the $U(1)_{DM}$ symmetry. 
The particle content and their charges under 
the $SM \otimes U(1)$ gauge groups are presented in tab.~\ref{tab:qunumU1}.
\begin{table}[h]
\begin{center}
\begin{tabular}{|c|c|c|c|c|}
\hline
Fields & $SU(3)_{c}$ & $SU(2)_{L}$ & $U(1)_{Y}$ & $U(1)_{DM}$ \\\hline
$L_{L}$ & 1 & 2 & -1/2 & 0 \\\hline
$l_{R}$ & 1 & 1 & -1 & 0 \\\hline
$H$   & 1  & 2 & 1/2  & 0 \\\hline
$N_{R,2,3}$ & 1 & 1 & 0 & 0 \\\hline
$\psi\equiv N_{R,1}$ & 1 & 1 & 0 & 3 \\\hline
$F=\left(\begin{matrix}
 F^{0} \\
 F^{-}
\end{matrix}\right)$  & 1  & 2 & -1/2 & 2 \\
$\overline{F}=\left(\begin{matrix}
 \overline{F}^{0} \\
 F^{+}
\end{matrix}\right)$  & 1  & $ \overline{2} $ & 1/2 & - 2 \\
\hline
$\Phi$   & 1  & 1 & 0  & 1\\\hline
\end{tabular}
\caption{The quantum numbers of Lepton, Higgs of SM and newly added fields under the SM as well as $U(1)_{DM}$ gauge groups.}
\label{tab:qunumU1}
\end{center}
\end{table}

The SM Lagrangian has to be extended with the following terms due to 
the addition of the new fields and extra $U(1)_{DM}$ gauge group.
\begin{itemize}
\item {\bf{The renormalizable terms :}}
\begin{eqnarray}
\mathcal{L}_{4}\,&=&\,\overline{F}i\D^{F}\,F+\overline{\psi}i\D^{\psi}\,\psi-M_{F}\overline{F}F-M_{\psi}\overline{\psi}\psi+\overline{N}_{R,i}i\Del\,N_{R,i}-\frac{M_{N,i}}{2}\overline{N}^{c}_{R,i}N_{R,i}
\nonumber\\
&&-Y_{\nu,i}\overline{N}_{R,i}L_{L}H +\left(D^{\Phi}_{\mu}\Phi\right)^{\dagger}(D^{\mu\,\Phi}\Phi)-\frac{m^{2}_{\Phi}}{2}\Phi^{\dagger}\Phi-\lambda_{H\Phi}\Phi^{\dagger}\Phi\,H^{\dagger}H-\lambda_{\Phi}(\Phi^{\dagger}\Phi)^{2}  \nonumber\\
& & -\frac{1}{4}B^{D}_{\mu\nu}B^{D\,\mu\nu}.
\label{eqn:dim4lag}
\end{eqnarray}
where $D^{F}_{\mu}=\partial_{\mu}-ig_{w}\dfrac{\tau^{a}}{2}W^{a}_{\mu}-\dfrac{1}{2}ig_YB_{Y\mu} -2ig_{D}Z_{D\mu}$,$\,D^{\psi}_{\mu}\,=\,\partial_{\mu}-3ig_{D}Z_{D\mu}$, $D^{\Phi}_{\mu}\,=\,\partial_{\mu}-ig_{D}Z_{D\mu}$ and 
$B^{D}_{\mu\nu}\,=\,\partial_{\mu}Z_{D\nu}-\partial_{\nu}Z_{D\mu}$.

\item {\bf{The dimension-5 terms : }}
\begin{eqnarray}
\mathcal{L}_{5}\,=\,&-&\frac{1}{M_{Pl}}\bigg[\overline{N}^{c}_{R,i}N_{R,i}\left(\Phi^{\dagger}\Phi+H^{\dagger}H\right)+\overline{F}F\left(\Phi^{\dagger}\Phi+H^{\dagger}H\right)+\overline{\psi}\psi\left(\Phi^{\dagger}\Phi+H^{\dagger}H\right)\nonumber\\
&&+(\overline{L}^{c}_{L}H)(L_{L}H)+(\overline{F}\sigma^{\mu\nu}F+\overline{\psi}\sigma^{\mu\nu}\psi)B^Y_{\mu\nu}\bigg]\nonumber\\
&-&\frac{f_{1}}{M_{Pl}}\left(\overline{F}H\,\psi\Phi^{\dagger}+h.c\right)-\frac{f_{2}}{M_{Pl}}\left(\overline{F}_{R}L_{L}\Phi^{2}+h.c\right).
\label{eqn:dim5lag}
\end{eqnarray}
The Wilson coefficients $f_{1}$ and $f_{2}$ can be $\ll 1$ if the  
corresponding dimension-5 terms are generated via some loop-induced 
diagrams at the Planck scale.
\end{itemize}

After the field $\Phi$ acquires a vacuum expectation value ($v_{D}$) 
the $U(1)_{DM}$ symmetry is spontaneously broken and the 
corresponding gauge boson called dark gauge-boson ($Z_{D}$) gets a 
mass $M_{Z_{D}}\,\sim\,g_{D}v_{D}$ and one can expand the Lagrangian 
with the replacement $\Phi\,=(v_{D}+S)/\sqrt{2}$ in the unitary 
gauge. Similarly one can replace 
$H\rightarrow\,(v+h)/\sqrt{2}$ in order 
to obtain possible mass terms and Higgs interaction terms. 
The lepton flavour violating decay $\mu\rightarrow e\gamma$ is mediated by $F^0-W^-$ loop or $F^--Z$ 
loop and the rate is dictated by the $L_{L}-F$ mixing angle $\theta_{-} \sim f_2 v^2_D/M_F M_{Pl}$. 
The present limit on the branching ratio of $\mu \rightarrow e\gamma$~\cite{TheMEG:2016wtm} puts a upper limit on $v_D$. 
On the other hand, the out-of-equillibrium condition on $\psi$ puts a lower bound on $v_D$ since the $Z_D$-boson can thermalize the
DM state via $2\rightarrow 2$ scatterings.  
We found $v_D = 7\times 10^{7}\,{\rm GeV}$  to be consistent with  both $\mu\rightarrow e\gamma$ as well as DM production.

The phenomenology of this model keeping $v_D = 7\times 10^{7}\,{\rm GeV}$ fixed, can briefly be stated as:
\begin{itemize}
\item {\bf{DM Production :}} The dominant contribution to the $\psi$ relic density originates also here from 
the decay $F \rightarrow\,\psi\, H$. But in this case, the interaction is generated by the dimension-5 effective 
operator  $\dfrac{f_{1}v_{D}}{M_{Pl}}\bar{F}H\,\psi$. In the right panel of fig.~\ref{fig:effsimpmod} we have shown the DM relic
density as a function of $f_1$ and DM mass, where we fix the mass of $F$ to be 1 TeV. The observed relic density can be achieved for 
$f_{1}\approx\,1$ and $v_{D}\,=7\,\times\,10^{7}\,$GeV.


 \begin{figure}
 \begin{center}
 \includegraphics[width=7cm,angle=0]{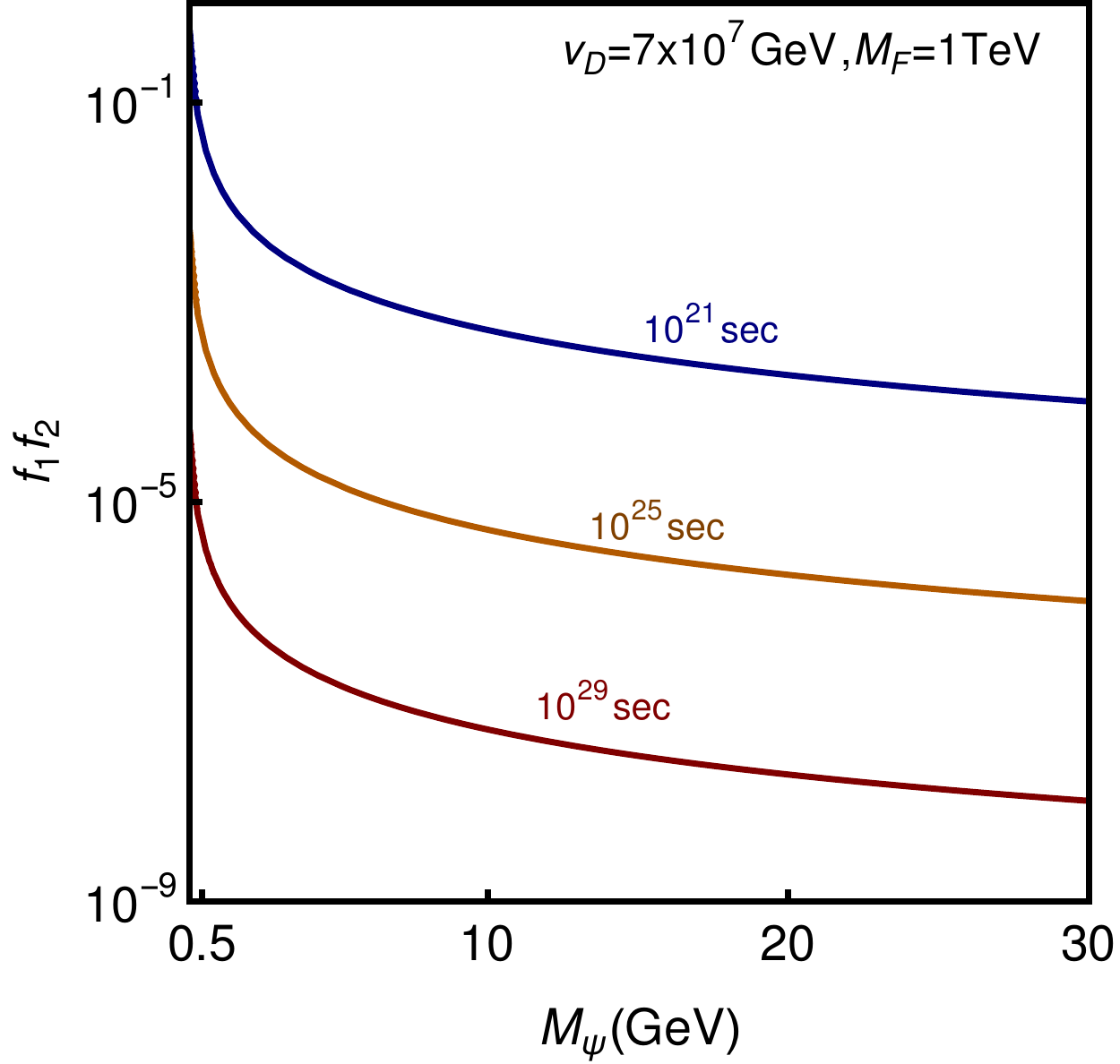}
  \includegraphics[width=7cm,angle=0]{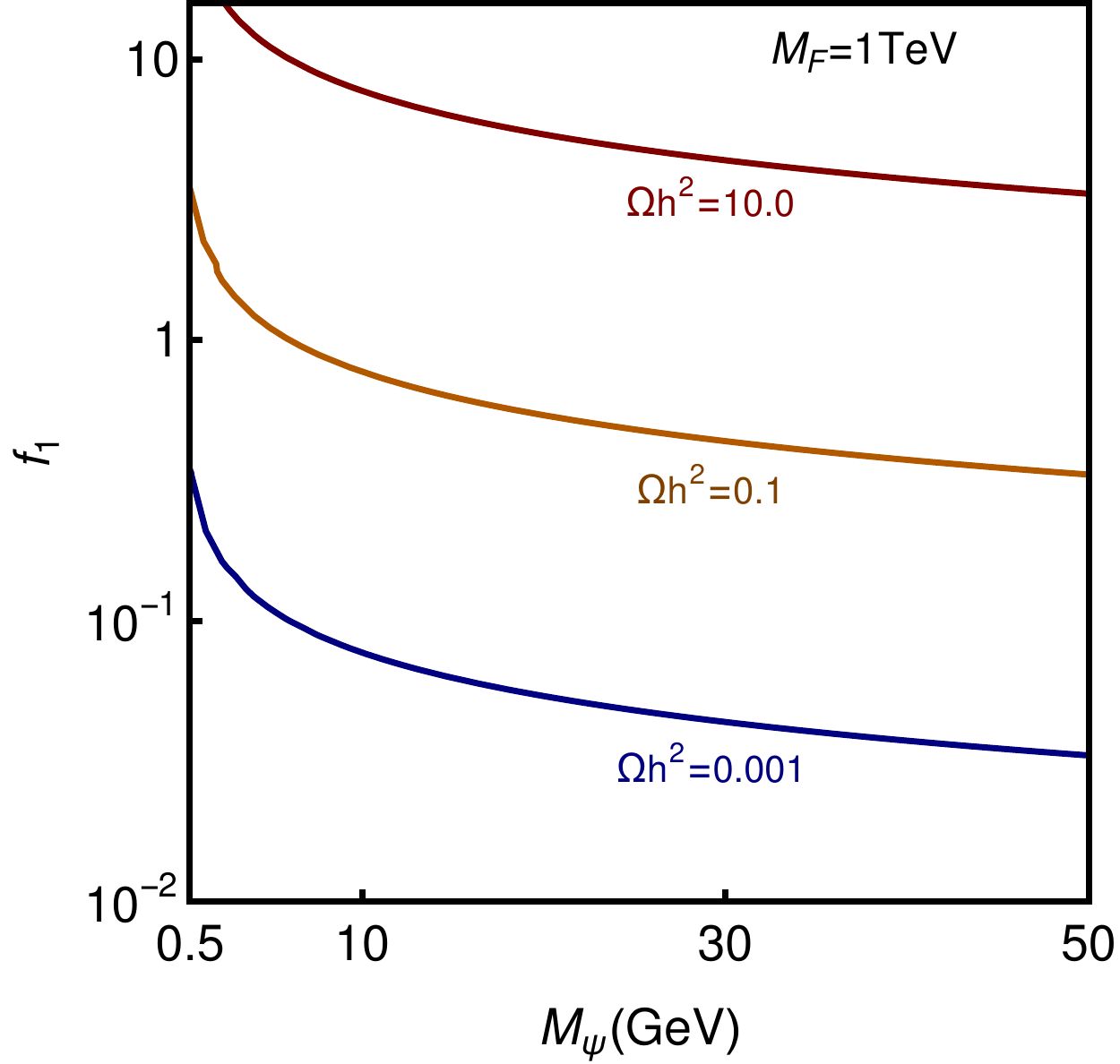}
 \caption{Left panel: DM lifetime contours in the $f_1 f_2$ vs. $M_\psi$ plane.
 Right panel: relic density contours in the $f_1-M_\psi$ plane.
 In both the cases we have considered $v_D=7\times 10^7\,{\rm GeV}$ and $M_F=1\,{\rm TeV}$. 
}
 \label{fig:effsimpmod}
 \end{center}
 \end{figure}

\item {\bf{DM decay :}}
The decay of DM $\psi$ into SM final states are taking place via intermediate off-shell fermions and are therefore driven 
by the factor $f_{1}f_{2}\,v\,v^{3}_{D}/(M^{2}_{Pl}\, M_\psi M_{F})$ with a combination of dimension 5 vertices. 
In the left panel of fig.~\ref{fig:effsimpmod} we have shown the DM lifetime contours in the plane of DM mass $M_\psi$ and 
Wilson coefficients $f_1 f_2$.  
For a chosen benchmark with $v_{D}\,=7\,\times\,10^{7}\,$GeV we found that the product of the two Wilson coefficients 
$f_1f_2$ has to be smaller than $\sim\,10^{-5}$ to achieve a dark matter lifetime of $10^{26}$~sec or more (see fig.~\ref{fig:effsimpmod}, \textit{left-panel}).
So in this case the suppressed decay can be achieved also for moderately small couplings.
Note that the $ f_1 $ coupling also drives the DM production and has to remain of order $ {\cal O} (1) $ for DM masses
in the tens of GeV, in order to produce a sufficient DM abundance, as shown in fig.~\ref{fig:effsimpmod}(\textit{right-panel}).

 \begin{figure}
 \begin{center}
  \includegraphics[width=7cm,angle=0]{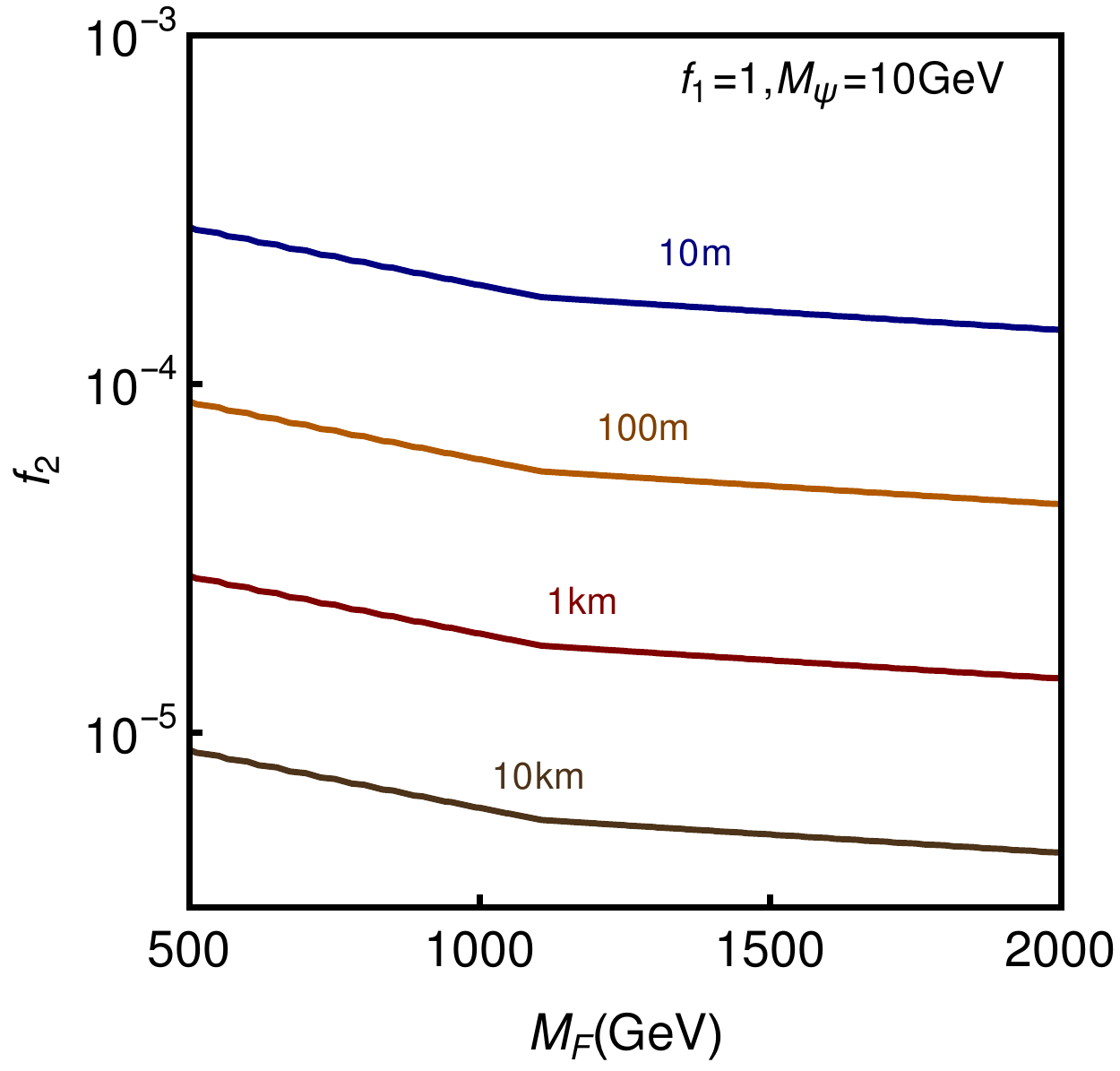}
 \caption{$F^0$ decay length contours in the $f_2-M_F$ plane. Here we have assumed $v_D=7\times 10^{7}\,{\rm GeV}$ and $M_\psi=10\,{\rm GeV}$. We have also chosen $f_1=1$ in order to have the $\textit{freeze-in}$ relic density in the correct ballpark.}
 \label{fig:effsimpmod2}
 \end{center}
 \end{figure}

\item {\bf{$F^{0}$ production and decay at colliders:}} 
The electroweak states $ F^{\pm}, F^0, \overline{F}^0 $ can be produced at colliders
via Drell-Yan production. Generically, we expect the charged states to be slightly heavier
than the neutral ones and be able to decay promptly into the neutral states and pions~\cite{Cirelli:2005uq}.
So a substantial population of $  F^0, \overline{F}^0 $ particles can arise even at the LHC,
if their mass is below $ 1 $ TeV.
The decay of $F^{0}$ occurs both in pure SM modes ($W(Z)l(\nu)$) or in a mixed
SM-BSM mode ($\psi\,h$). The latter mode contributes to $\textit{freeze-in}$ relic 
density of $\psi$ and hence has to be very slow while the decay into 
SM states can occur faster via the mixing $f_{2}v^{2}_{D}/M_{Pl}M_{F}$. 
The dominant part of the decay width of $F^{0}$ is therefore given by,
\begin{equation}
\Gamma_{F^{0}\rightarrow\,W(Z)l(\nu)}\,=\,\frac{g^{2}_{weak}|V_{\nu\,F}|^{2}(c^{2}_{V}+c^{2}_{A})}{8\pi}\frac{M^{3}_{F}}{M^{2}_{V}}\left(1-\frac{M^{2}_{V}}{M^{2}_{F}}\right)^{2}\left(1+\frac{2\,M^{2}_{V}}{M^{2}_{F}}\right).
\end{equation}
where $V_{\nu F}$ is proportional to $f_{2}v^{2}_{D}/M_{Pl}M_{F}$.
Depending on the value of $f_2$,  $F^{0}$ can have a decay length of few meters to one kilometer as shown
in fig.~\ref{fig:effsimpmod2}. So the mother particle in this scenario can realised both displaced vertices or
missing energy signatures and can be searched at present and future colliders \cite{Arcadi:2014tsa, Acharya:2014nyr,Curtin:2018mvb}.

Of course we could also lower the scale of the non-renormalizable operators from $ M_{Pl} $ to some intermediate
scale $ \Lambda $  and obtain still a consistent picture, as long as we satisfy
$\dfrac{f_{1}v_{D}}{\Lambda}\sim 10^{-12} $ and the DM lifetime remains sufficiently long.
Note that while the production is driven by the factor $ \dfrac{v_D}{\Lambda} $, the decay depends on the
combination $  \dfrac{ f_1 f_2 v v_D^3}{\Lambda^2 M_\psi M_F} \leq 10^{-24}  \dfrac{ f_2 v v_D}{ f_1 M_\psi M_F}$
assuming the value of the effective coupling from the FIMP production.
Then for a dark matter mass of $10 $ GeV, this gives a lower bound on the value of $v_D $ as
\begin{equation}
v_D \leq 62\; \frac{f_1}{f_2} \; M_F  
\end{equation}
We see therefore that in that case the dark Sector could be characterised by a similar mass scale for the
scalar and fermionic states and the UV completion of the model may appear way below the Planck scale.

\end{itemize}
The overall advantages of this framework are:
\begin{itemize}
\item The effective operators in eq.~(\ref{eqn:dim5lag}), suppressed by the Planck mass, successfully generate 
the effective couplings involved in DM decay as well as $\textit{freeze-in}$ in the right ballpark, without the
need to fine-tune the Wilson coefficients.  Thus a rather tantalizing connection with UV completion at the Planck scale arises,
even if also lower values of the cut-off scale are possible.

\item The scenario is cosmologically consistent and anomaly free.


\item The generation of neutrino masses and as well as various constraints from electroweak phenomenology  are not affected. 

\item The addition of the $U(1)_D $ breaking scalar $\Phi$ with the mass 
scale it brings in, enables one to achieve vacuum stability all the way to Planck scale~\cite{EliasMiro:2012ay}. 
\end{itemize} 

 On the other hand, there is no direct connection between neutrino phenomenology and DM phenomenology. 
 It is straight-forward to understand that such a connection can readily be established if we elevate the 
 $U(1)_{\rm DM}$ to $U(1)_{L_\mu-L_\tau}$~\cite{Baek:2008nz,Baek:2015fea,Patra:2016shz,Altmannshofer:2016jzy,Biswas:2016yjr,Biswas:2016yan,Biswas:2017ait} 
 or $U(1)_{B-L}$~\cite{Okada:2010wd,Lindner:2011it,Okada:2012sg,Basso:2012ti,Basak:2013cga,Sanchez-Vega:2014rka,Guo:2015lxa,Rodejohann:2015lca,Patra:2016ofq,Biswas:2017tce,Okada:2018ktp,Bhattacharya:2019tqq}. 
 We consider $U(1)_{L_\mu-L_\tau}$ in the next model.

\section{Gauged $U(1)_{L_\mu-L_\tau}$ Model} \label{sec:mutaumodel}

Now we will move to the gauged $U(1)_{L_\mu-L_\tau}$ model. 
This model can explain neutrino mass and give a decaying FIMP dark matter without any ad-hoc symmetry barring the 
$U(1)_{L_\mu-L_\tau}$. As has been mentioned in the introduction, such a scenario can help in identifying a common UV completion of 
the modeling of slowly decaying DM and the observed pattern in the neutrino sector, the degree of oscillation required to explain the data on atmospheric neutrinos. In addition to the SM particle content we consider again a symmetry breaking
scalar $\Phi$ and a vector-like fermion $\psi$, playing the role of the DM. So in this case the dark matter particle is not one of the heavy neutrinos, but it is still tightly
related to the leptonic sector via the gauge symmetry and the gauge symmetry preserving interactions.

The particle content of our model and their charges under all the gauge groups are shown in tab.~\ref{tab:qunumUlmultau}.
\begin{table}[h]
\begin{center}
\begin{tabular}{|c|c|c|c|c|c|}
\hline
Fields & Spin & $SU(3)_{c}$ & $SU(2)_{L}$ & $U(1)_{Y}$ & $U(1)_{L_\mu - L_\tau}$ \\\hline
$L_{L\,e}$ & 1/2 & 1 & 2 & -1/2 & 0 \\
$l_{R\,e}$ & 1/2 & 1 & 1 & -1 & 0 \\
$N_{R\,e}$ & 1/2 & 1 & 1 & 0 & 0 \\\hline
$L_{L\,\mu}$ & 1/2 & 1 & 2 & -1/2 & 1 \\
$l_{R\,\mu}$ & 1/2 & 1 & 1 & -1 & 1 \\
$N_{R\,\mu}$ & 1/2 & 1 & 1 & 0 & 1 \\\hline
$L_{L\,\tau}$ & 1/2 & 1 & 2 & -1/2 & -1 \\
$l_{R\,\tau}$ & 1/2 & 1 & 1 & -1 & -1 \\
$N_{R\,\tau}$ & 1/2 & 1 & 1 & 0 & -1 \\\hline
$\Phi$   & 0 & 1  & 1 & 0  & 1\\\hline
$\psi_{L,R}$ & 1/2 & 1 & 1 & 0 & 4 \\\hline
\end{tabular}
\caption{The quantum numbers of the SM Leptons and added BSM fields under SM as well as $U(1)_{L_\mu-L_\tau}$ gauge group.}
\label{tab:qunumUlmultau}
\end{center}
\end{table}

Apart from the tree level terms we have also considered possible higher order operators and the Lagrangian of the model consist of three pieces:
\begin{equation}
\mathcal{L} = \mathcal{L}_{dim-4} + \mathcal{L}_{dim-5} + \mathcal{L}_{dim-6}.
\end{equation} 
The dim-4 terms are given by,
\begin{eqnarray}
\mathcal{L}_{dim-4} &\supset & \overline{N}_{R i}i\D^{N,i}N_{R i}-\frac{1}{2}M_{ee}\overline{N}^c_{R e}N_{R e} - \frac{1}{2}M_{\mu\tau}\left(\overline{N}^c_{R\mu}N_{R\tau}+\overline{N}^c_{R\tau}N_{R\mu}\right)\nonumber\\
&&  -\frac{1}{2}h_{e\mu}\left(\overline{N}^c_{R e}N_{R\mu}+\overline{N}^c_{R\mu}N_{R e}\right)\Phi^\dagger -\frac{1}{2}h_{e\tau}\left(\overline{N}^c_{R e}N_{R\tau}+\overline{N}^c_{R\tau}N_{R e}\right)\Phi \nonumber\\
&&+\overline{\psi}i\D^{\psi}\psi-M_{\psi}\overline{\psi}{\psi} -\underset{\alpha=e,\mu,\tau}{\sum}Y_\alpha \overline{L}_{L \alpha} N_{R\alpha}H + V(\Phi,H)\\
&&-\dfrac{1}{4}B^{D}_{\mu\nu}B^{D\mu\nu}-\dfrac{\sin\epsilon}{2}B^{Y}_{\mu\nu}B^{D\mu\nu},
\label{eqn:Ldim4U1lmultau} 
\end{eqnarray}
with the scalar potential, 
\begin{equation}
V(\Phi,H) = -m^2_{\Phi}\Phi^\dagger \Phi +\lambda_{\Phi}(\Phi^\dagger \Phi)^2 +  \lambda_{H\,\Phi}\; \Phi^\dagger \Phi\; H^\dagger H.
\end{equation}
The presence of the term $\lambda_{H\Phi} \Phi^\dagger \Phi H^\dagger H$ causes mixing between the scalars $h$ and $\phi$ when the respective scalar 
fields acquire v.e.vs $v$ and $v_D$ respectively. We have minimized the potential $V(\Phi,H)$ to obtain,
\begin{eqnarray}
v\,=\,\left(\frac{2m^{2}_{H}\lambda_{\Phi}-\lambda_{H\Phi}m^{2}_{\Phi}}{\lambda^{2}_{H\Phi}-4\lambda_{H}\lambda_{\Phi}}\right)^{1/2}\hspace{0.5cm} {\rm and}\hspace{0.5cm}
v_{D}\,=\,\left(\frac{2m^{2}_{\Phi}\lambda_{H}-\lambda_{H\Phi}m^{2}_{H}}{\lambda^{2}_{H\Phi}-4\lambda_{H}\lambda_{\Phi}}\right)^{1/2}.
\label{eqn:vacexpval}
\end{eqnarray} 
Diagonalizing the mass matrix give rise to the mass eigenstates:
\begin{eqnarray}
\begin{pmatrix}
 h_1\\h_2
\end{pmatrix}
=
\begin{pmatrix}
 \cos\theta_h & \sin\theta_h \\ -\sin\theta_h & \cos\theta_h
\end{pmatrix}
\begin{pmatrix}
 h\\\phi
\end{pmatrix}
\end{eqnarray}
where the mixing angle $\theta_h \approx \dfrac{ \lambda_{H\Phi} v v_D}{(m^2_\Phi - m^2_h)+\frac{v^2}{2}(\lambda_{H\Phi}-\lambda_H)+\frac{v^2_D}{2}(\lambda_{\Phi}-\lambda_{H\Phi} ) }$.  

The kinetic mixing term $\dfrac{\sin\epsilon}{2}B^{Y}_{\mu\nu}B^{D\mu\nu}$ in the Lagrangian $\mathcal{L}_{dim-4}$ will induce a mixing between the 
SM-$Z$ boson and $L_\mu-L_\tau$ gauge boson $Z_D$. The mixing angle is given by~\cite{Arcadi:2018tly},
\begin{equation}
\tan\,2\alpha = -\frac{\hat{m}^2_Z \sin\theta_W \sin\,2\epsilon}{\hat{M}^2_{Z_D}- \hat{m}^2_Z (\cos^2\epsilon-\sin^2\epsilon \sin^2\theta_W) }\; ,
\label{eqn:kinmixZzp}
\end{equation}
where $\hat{m}_Z$, $\hat{M}_{Z_D}$ are the bare masses for the SM-Z boson and $Z_D$ boson respectively. Clearly the mixing angle $\alpha$ is strongly suppressed if the dark gauge boson $ Z_D $ is much heavier than the $Z$.

This mixing in turn induces a $\bar{\psi}\psi\,Z$ coupling $g_{\bar{\psi}\psi\,Z}=\dfrac{\sin\alpha}{\cos\epsilon}g_D$ and modifies the 
lepton couplings to SM $Z$ boson:
\begin{eqnarray}
g_{f_L\,Z}&=&\frac{e}{\sin\theta_W\,\cos\theta_W}\cos\alpha\left[T_3\left(1+\sin\theta_W\,\tan\epsilon\tan\alpha\right)-Q_f\left(\sin^2\theta_W+\sin\theta_W\tan\epsilon\tan\alpha\right)\right]\nonumber\\
&&\pm\,g_D \frac{\sin\alpha}{\cos\epsilon}\\
g_{f_R\,Z}&=&-\frac{e}{\sin\theta_W\,\cos\theta_W}\cos\alpha\,Q_f\left(\sin^2\theta_W + \sin\theta_W\tan\epsilon\tan\alpha\right)\pm\,g_D \frac{\sin\alpha}{\cos\epsilon},
\end{eqnarray}
where $\left(+g_D \frac{\sin\alpha}{\cos\epsilon}\right)$ is relevant for $f=\mu$ and $\left(-g_D \frac{\sin\alpha}{\cos\epsilon}\right)$ is for $f=\tau$. 
The first term in each case is the corresponding coupling for other SM fermions.  

For our choice of parameters $g_D=0.01,v_D \geq 10^7\,{\rm GeV}$ we get $\hat{M}_{Z_D} \geq10^5\,{\rm GeV}$ and hence 
$\sin\alpha \leq 10^{-7}$ even for $\sin\epsilon \approx 1$. 
This happens because the mixing angle $\alpha$ is proportional to $\hat{m}^2_Z/\hat{M}^2_{Z_D}$ which is always small due to large $v_D$.
This causes a $\bar{\psi}\psi\,Z$ coupling $\leq \mathcal{O}(10^{-8})$ and thus one can neglect the effect of kinetic mixing in the following analysis. 

The dimension-5 Lagrangian consists of the following terms,
\begin{eqnarray}
\mathcal{L}_{dim-5} &\supset & \frac{f_1}{\Lambda}\overline{\psi}\psi \left(\Phi^\dagger \Phi + H^\dagger H \right) 
+ \frac{f_2}{\Lambda}\;  \overline{N}^c_{R e}N_{R e}\Phi^\dagger \Phi \nonumber\\
&&+\frac{f_3}{\Lambda} \left(\overline{N}^c_{R\mu}N_{R\mu}\Phi^{\dagger 2} + h.c \right) +\frac{f_{3^\prime}}{\Lambda} \left( \overline{N}^c_{R\tau}N_{R\tau}\Phi^{2} + h.c \right)   + \frac{f_5}{\Lambda}   \underset{\alpha=e,\mu,\tau}{\sum} \overline{L}^c_{L \alpha} H \overline{L}_{L \alpha} H\nonumber\\
&& +\frac{f_6}{\Lambda} H\left(\overline{L}_{Le} N_{R\mu}\Phi^\dagger +\overline{L}_{L\mu} N_{R e}\Phi \right)+\frac{f_7}{\Lambda} H \left(\overline{L}_{Le}N_{R\tau}\Phi + \overline{L}_{L\tau} N_{R e}\Phi^\dagger \right)
\label{eqn:Ldim5U1lmultau} 
\end{eqnarray}
There is no  interaction between DM $\psi$ and other fields due to $L_\mu-L_\tau$ charge of the DM at 
this level but these terms have an important role to play in the neutral lepton mass matrix and thus in the $\psi-N$ mixing. 
The dimension-6 terms $\mathcal{L}_{dim-6}$ induce the lowest order mixing between DM $\psi$ 
and SM sector\footnote{We have listed only the terms which affects the dark matter and neutrino phenomenology we are going to study hereafter.}:
\begin{eqnarray}
\mathcal{L}_{dim-6} &\supset & \frac{f_4}{\Lambda^2}\left[\left(\overline{\psi}_L N_{R\mu}+\overline{\psi}_R N^c_{R\tau} \right)\Phi^3 + h.c \right] + \frac{f_8}{\Lambda^2}H\left(\overline{L}_{L\mu}N_{R\tau}\Phi^{\dagger 2}+\overline{L}_{L\tau}N_{R\mu}\Phi^2 \right).\nonumber\\
\label{eqn:Ldim6U1lmultau} 
\end{eqnarray}
Note that here we consider a generic cut-off scale $ \Lambda \leq M_{Pl} $ and contrary to the previous case, we consider as well
the dimension 6 operators, as  there is no mixing between the neutrinos and the dark matter from the lower order operators.

\subsection{Neutrino Phenomenology} 
\label{sec:neutrino}

The neutral lepton mass terms after $U(1)_{L_\mu - L_\tau}$ breaking takes the form $V^\dagger \mathcal{M}_0 V$, 
where $V=(\nu^c_{L,\alpha}, N_{R,\alpha}, \Psi_R)^T$ with $\alpha=e,\mu,\tau$ and $\Psi_R = (\psi^c_L, \psi_R)$.  The mass matrix is,
\begin{equation}
\mathcal{M}_0 =  \left(\begin{matrix}
0 & \dfrac{Y_\alpha v}{\sqrt{2}} & 0\\\vspace{1mm}
\dfrac{Y^T_\alpha v}{\sqrt{2}} & \mathcal{M}_{N} & \dfrac{Y_{N\psi} v^3_D}{\Lambda^2} \\\vspace{1mm}
0 & \dfrac{Y^T_{N\psi} v^3_D}{\Lambda^2} & \mathcal{M}_\Psi \end{matrix}\right).
\label{eqn:massmat}
\end{equation}
where, 
\begin{eqnarray}
Y_\alpha\,&=&\,\left(\begin{matrix}
Y_e & \dfrac{f_6 v_D}{\Lambda} & \dfrac{f_7 v_D}{\Lambda} \\\vspace{2mm}
\dfrac{f_6 v_D}{\Lambda} & Y_\mu& \dfrac{f_8 v^2_D}{\Lambda^2} \\\vspace{2mm}
\dfrac{f_7 v_D}{\Lambda} & \dfrac{f_8 v^2_D}{\Lambda^2}  & Y_\tau  \end{matrix}\right),\hspace{1cm}
Y_{N\psi}\,=\,\left(\begin{matrix}
0 & 0  \\\vspace{1mm}
f_4 & 0 \\\vspace{1mm}
0 & f_4 \end{matrix}\right),
\end{eqnarray}\\
\begin{eqnarray}\label{eq:MN}
\mathcal{M}_{N}\,&=&\,\left(\begin{matrix}
\frac{1}{2}M_{ee} & h_{e\mu} v_D & h_{e\tau} v_D \\\vspace{1mm}
h^T_{e\mu} v_D & \dfrac{f_3 v^2_D}{\Lambda} & M_{\mu\tau}\\\vspace{1mm}
h^T_{e\tau} v_D & M_{\mu\tau} &  \dfrac{f_3 v^2_D}{\Lambda} \end{matrix}\right),\hspace{1cm}
\mathcal{M}_{\Psi}\,=\,\left(\begin{matrix}
0 & M_{\psi}  \\\vspace{1mm}
M_{\psi}  & 0 \end{matrix}\right). 
\end{eqnarray}
For simplicity we neglect $f_8 v^2_D/\Lambda^2$ since this term is suppressed by $v_D/\Lambda$ compared to the terms proportional 
to $f_{6,7}$, as well as the terms $ {\cal O} (v_D^3/\Lambda^2) $ in $\mathcal{M}_0$.  
Diagonalizing the mass matrix $\mathcal{M}_0$ we obtain then,
\begin{eqnarray}
\mathcal{M}_N &=& \left(\frac{Y_\alpha v}{\sqrt{2}}\right)U_{PMNS}\left(m^{diag}_\nu
\right)^{-1}U^T_{PMNS}\left(\frac{Y_
\alpha v}{\sqrt{2}}\right)^{T}. 
\label{eqn:matrieqnnew}
\end{eqnarray}
Here $m^{diag}_\nu = \textrm{diag}(m_1,m_2,m_3)$, where $m_1$ is mass of the lightest neutrino.
Comparing the (22) and (33) elements of the above equation with $\mathcal{M}_{N}$ in eq.~(\ref{eq:MN}) we get,
\begin{eqnarray}
&&\left(\frac{f_6 v_D}{\Lambda}\right)^2 \left(\frac{U^2_{e1}}{m_1} + \frac{U^2_{e2}}{m_2} + \frac{U^2_{e3}}{m_3}\right)+ 2\left(\frac{f_6 v_D}{\Lambda}Y_\mu\right)\left(\frac{U_{e1}U_{\mu 1}}{m_1}+\frac{U_{e2}U_{\mu 2}}{m_2}+\frac{U_{e3}U_{\mu 3}}{m_3}\right)\nonumber\\
&&\hspace{6.2cm}+Y^2_\mu\left(\frac{U^2_{\mu 1}}{m_1}+\frac{U^2_{\mu 2}}{m_2}+\frac{U^2_{\mu 3}}{m_3} \right)=\frac{2}{v^2}\frac{f_3 v^2_D}{\Lambda}
\label{eqn:mastereq1}
\end{eqnarray}
and,
\begin{eqnarray}
&&\left(\frac{f_7 v_D}{\Lambda}\right)^2 \left(\frac{U^2_{e1}}{m_1} + \frac{U^2_{e2}}{m_2} + \frac{U^2_{e3}}{m_3}\right)+ 2\left(\frac{f_7 v_D}{\Lambda}Y_\tau\right)\left(\frac{U_{e1}U_{\tau 1}}{m_1} + \frac{U_{e2}U_{\tau 3}}{m_2} + \frac{U_{e3}U_{\tau 3}}{m_3}\right)\nonumber\\
&&\hspace{6.2cm}+Y^2_\tau\left( \frac{U^2_{\tau 1}}{m_1} + \frac{U^2_{\tau 2}}{m_2} + \frac{U^2_{\tau 3}}{m_3} \right)=\frac{2}{v^2}\frac{f_{3^\prime} v^2_D}{\Lambda}.
\label{eqn:mastereq2}
\end{eqnarray}
Here $U_{\alpha i}$ are elements of the $U_{PMNS}$ matrix:
\begin{eqnarray}
U_{PMNS} &\equiv & \left(
\begin{matrix}
c_{12}c_{13} & s_{12}c_{13} & s_{13}e^{-i\delta_{\rm CP}} \\
-s_{12}c_{23}-c_{12}s_{23}s_{13}e^{i\delta_{\rm CP}} & (c_{12}c_{23}-s_{12}s_{23}s_{13}e^{i\delta_{\rm CP}}) & s_{23}c_{13} \\
s_{12}s_{23}-c_{12}c_{23}s_{13}e^{i\delta_{\rm CP}} & (-c_{12}s_{23}-s_{12}c_{23}s_{13}e^{i\delta_{\rm CP}}) & c_{23}c_{13}
\end{matrix}\right) \times \mathcal{P}
\nonumber \\
\end{eqnarray}
with $\mathcal{P}\,=\,{\rm diag}(1,e^{i\alpha/2},e^{i\beta/2} )$ and  $c_{12}=\cos\theta_{12}$ etc. The eqs.~(\ref{eqn:mastereq1}) and (\ref{eqn:mastereq2}) essentially determine the parameter space of the model. 
Interestingly, in the usual $L_\mu-L_\tau$ scenario one does not consider the higher-dimensional terms and  it becomes difficult to obtain 
a neutrino mass spectra consistent with the PLANCK constraint of $\underset{i=1,2,3}{\sum} |m_{\nu i}| \leq 0.11\,{\rm eV}$~\cite{Aghanim:2018eyx}. 
In our case instead we obtain a large parameter space satisfying neutrino oscillation data which also satisfy the PLANCK constraint as we will discuss below.

By diagonalizing the mass matrix $\mathcal{M}_0$ one also obtains the mixing between DM $\psi$ and SM neutrinos which is given by, 
\begin{equation}
U_{\nu\psi} \approx -\frac{Y^T_{N\Psi}Y^T_\alpha}{\mathcal{M}_N \mathcal{M}_\Psi}\left(\frac{v v^3_D}{\sqrt{2} \Lambda^2}\right).
\label{eqn:DMmix}
\end{equation}
Following eqs.~(\ref{eqn:mastereq1}), (\ref{eqn:mastereq2}) and (\ref{eqn:DMmix}) one can understand that $v_D/\Lambda $ and
$ v_D^2/\Lambda $ are again the key parameters and they determine now both the neutrino as well as the DM phenomenology.
One can further see from eq.~(\ref{eqn:DMmix}) that the DM-SM neutrino mixing angle $U_{\nu\psi}$ is not only determined by 
the combination $v^3_D/\Lambda^2$ but also by the sterile neutrino-active neutrino mixing angle $U_{\nu N} \simeq Y^T_\alpha v/\sqrt{2}\mathcal{M}_N $ 
and thus the DM phenomenology is quite entangled with the neutrino phenomenology in this scenario.

Let us first consider the neutrino sector and  eqs.~(\ref{eqn:mastereq1}) and (\ref{eqn:mastereq2}).
A large number of parameters appears there, and to understand
the dependence on them one need to assume only some of them dominate in the equations. 
For example, if we consider the case 
$v_D/\Lambda \gg Y_{\mu,\tau}$ then only the first terms in  eqs.~(\ref{eqn:mastereq1}) and (\ref{eqn:mastereq2}) are important 
and under the assumption of $f_{6,7,3,3^\prime} \approx 1$, one needs
$\Lambda \simeq \dfrac{v^2}{2}\left(\frac{U^2_{e1}}{m_1} + \frac{U^2_{e2}}{m_2} + \frac{U^2_{e3}}{m_3}\right) \sim 10^{14}\,{\rm GeV}$,
similar to the scale needed by generating the light neutrino masses with the dimension-5 Weinberg operator. 
In the opposite extreme, taking $v_D/\Lambda \ll Y_{\mu,\tau}$,  one obtains instead
$Y^2_i \simeq \dfrac{2}{v^2} \dfrac{f_{3,3^\prime}v^2_D}{\Lambda} \left( \frac{U^2_{i 1}}{m_1} + \frac{U^2_{i 2}}{m_2} + \frac{U^2_{i 3}}{m_3} \right)^{-1} \simeq 10^{-14}\,{\rm GeV}^{-1}\,\dfrac{f_{3,3^\prime}v^2_D}{\Lambda} $ (here $i=\mu,\tau$).
Now it is clear that if $f_{3,3^\prime} v^2_D/\Lambda \lesssim 10^{-2}$ then the Majorana neutrino  masses are of $\mathcal{O}(100{\rm MeV})$ and will have very suppressed mixing angles $\theta_{\nu N} \sim 10^{-5} $.  These neutrinos will therefore be extremely long-lived  and their late time decays will be in tension with BBN prediction of light-element abundances. 
A large value of $v^2_D/\Lambda$ will decrease the DM lifetime (following eq.~(\ref{eqn:DMmix})) and we will need to push $f_4$ to smaller values to make DM stable till today. Therefore only a window of values for $ v^2_D/\Lambda $ is viable, where the lower bound is set by the neutrino sector and upper bound is dictated by the DM lifetime.

\begin{figure}[t!]
\begin{center}
\includegraphics[width=7.5cm]{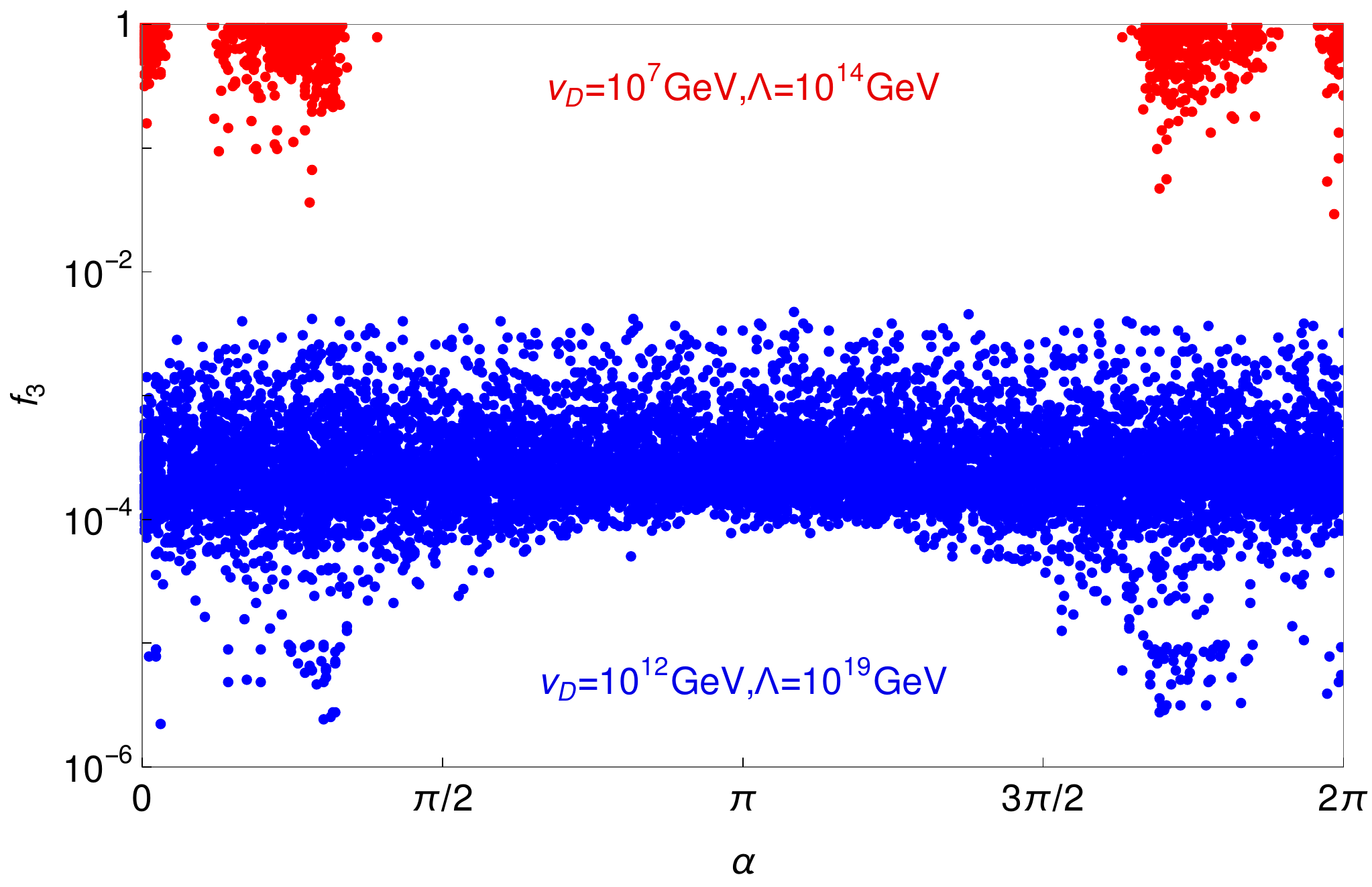}
\includegraphics[width=7.5cm]{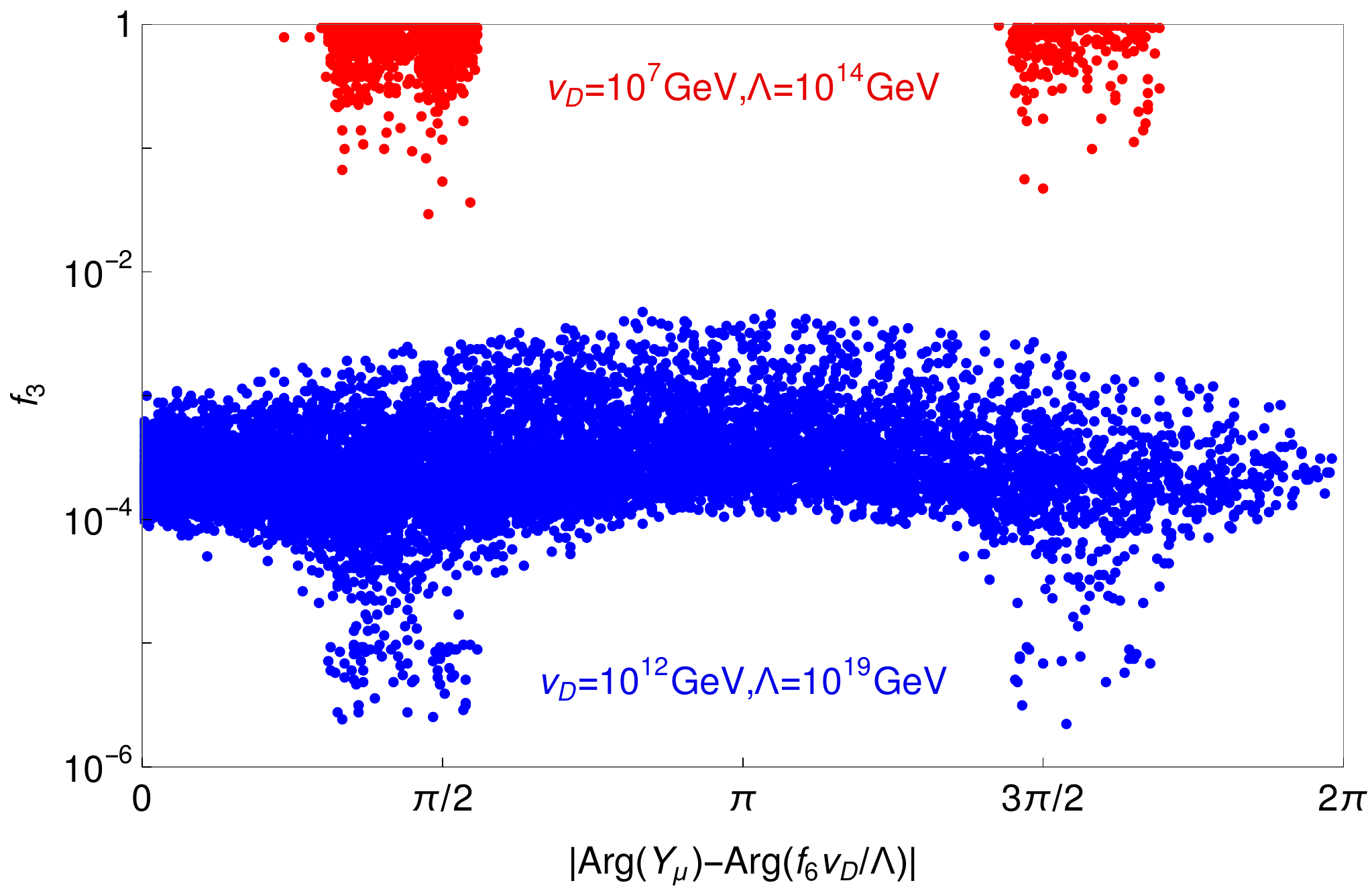}
\end{center}
\caption{Left Panel: Value of the coupling $ f_3 $ giving a solution of the condition in eq.~(\ref{eqn:mastereq1}) as a function of the 
Majorana phase $ \alpha $ in the PMNS mixing matrix.
Right Panel: Value of the coupling $ f_3 $ giving a solution of the condition in eq.~(\ref{eqn:mastereq1}) as a function of the 
Yukawa and Wilson coefficients phases.
For smaller values of $ v_D $, a correlation among the phases becomes apparent, which disappears for larger values.
}
\label{fig:neutrinof3vsalpha}
\end{figure}

The intermediate situation $Y_{\mu,\tau} \simeq v_D/\Lambda$  is more interesting to look at, as then all the terms in the {\it l.h.s}
are important. Under the assumption of $Y_{\mu,\tau} \simeq v_D/\Lambda$ with a scan over $v_D$ and $\Lambda$ we did not find any parameter point 
consistent with $|f_3|,|f_{3^\prime}| \leq 1$ for $v_D \lesssim 10^7\,{\rm GeV}$ and $\Lambda \lesssim 10^{14}\,{\rm GeV}$. 
This is mainly because keeping $Y_{\mu,\tau} \simeq v_D/\Lambda$ fixed if one reduces the values of $v_D$ and $\Lambda$, $v^2_D/\Lambda$ 
becomes too small and thus it is not possible to satisfy $|f_3|,|f_{3^\prime}| \leq 1$. We have chosen $v_D=10^7\,{\rm GeV}$, $\Lambda=10^{14}\,{\rm GeV}$ 
as our benchmark in the following analysis. 
For larger values of $ v_D, \Lambda $ with the same ratio $Y_{\mu,\tau} \simeq v_D/\Lambda $, the couplings $|f_3|,|f_{3^\prime}| $ can be
very small and the eqs.~(\ref{eqn:mastereq1}) and (\ref{eqn:mastereq2}) are satisfied without any cancellation among different terms in the {\it  l.h.s.}. 
As an illustration we have shown in fig.~\ref{fig:neutrinof3vsalpha} the allowed values of $\alpha$ and $|Arg(Y_{\mu})-Arg(f_6v_D/\Lambda)|$ for 
two different benchmarks $(v_D,\Lambda)=(10^7\,{\rm GeV},10^{14}\,{\rm GeV})$ and $(10^{12}\,{\rm GeV},10^{19}\,{\rm GeV})$. 
Although the values of $v_D/\Lambda \simeq 10^{-7}$ in both the cases, $v^2_D/\Lambda$ is larger in the latter scenario. 
Thus the distribution of parameter points for $(v_D,\Lambda)=(10^7\,{\rm GeV},10^{14}\,{\rm GeV})$(red-dotted) is shifted upward by a 
factor of $10^5$ compared to the case  $(v_D,\Lambda)=(10^{12}\,{\rm GeV},10^{19}\,{\rm GeV})$(blue-dotted).
Thus we see that for $(v_D,\Lambda)=(10^7\,{\rm GeV},10^{14}\,{\rm GeV})$ one needs cancellation among several terms in 
the \textit{l.h.s} of eqs.~(\ref{eqn:mastereq1}) and (\ref{eqn:mastereq2}). Such a cancellation occurs if 
$|Arg\left(Y_{\mu(\tau)}\right)-Arg\left(f_{6(7)}v_D/\Lambda\right)|\,\simeq \pi/2$ or $3\pi/2$ and $\alpha \simeq 0\,$or $\,2\pi$. 

\begin{table}[t]
\begin{center}
\begin{tabular}{|c|c||c|c|}
\hline
Fixed Parameters & Values & Varied Parameters & Ranges\\\hline
$v_D$ & $10^7$ GeV & $m_{1}$ &  $[0.001\,,0.026]\,{\rm eV}$\\
$\Lambda$ & $10^{14}$ GeV & $\delta_{CP}$ & $[0,2\pi]$\\\hline
$|f_{6,7}|$ & 1 & $\alpha$ & $[0,2\pi]$ \\
$|Y_{e,\mu,\tau}|$ & $10^{-7}$ & $\beta$ & $[0,2\pi]$ \\\hline
$\theta_{12}$ & 33.85$^o$ &  Arg($f_6$) & $[0,2\pi]$ \\
$\theta_{23}$ & 48.35$^o$ & Arg($Y_\mu$) & $[0,2\pi]$ \\
$\theta_{13}$ & 8.61$^o$ & Arg($f_7$) & $[0,2\pi]$ \\
 & &  Arg($Y_\tau$) & $[0,2\pi]$ \\\hline
\end{tabular}
\caption{Parameters kept fixed during our analysis for producing fig.~\ref{fig:neutrinopredictsplot} and the parameters which have been varied within certain ranges are tabulated.}
\label{tab:benchmarkLmuLtau}
\end{center}
\end{table}

\begin{figure}[t!]
\begin{center}
\includegraphics[width=12cm]{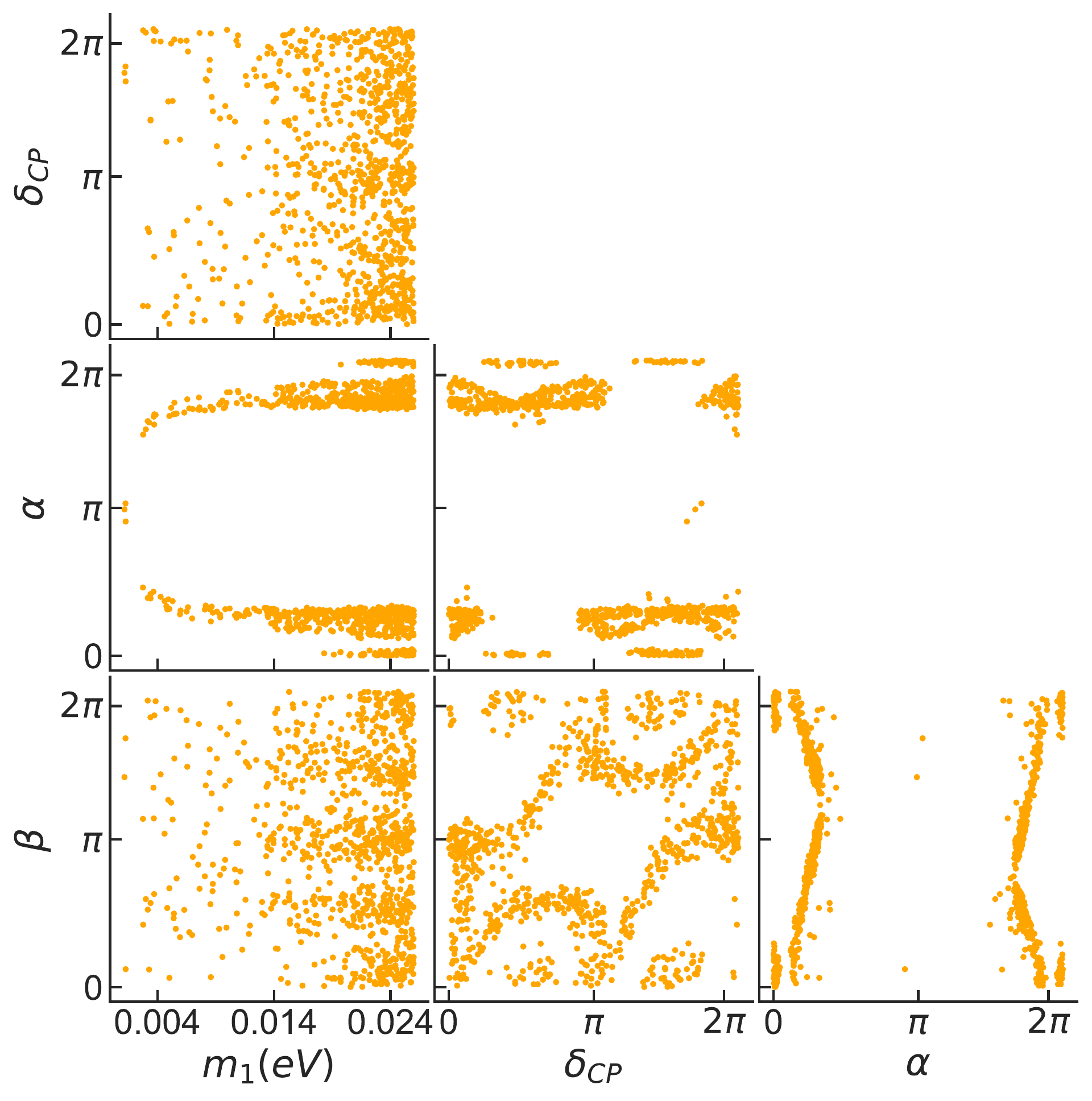}
\end{center}
\caption{Correlation among different phases relevant for neutrino oscillation: the CP-phase($\delta_{\rm CP}$) and the Majorana phases ($\alpha,\beta$). The correlation of each of these phases with the lightest neutrino mass($m_1$) is shown in the left most column.  These correlation is found under the assumption of tab.~\ref{tab:benchmarkLmuLtau}. }
\label{fig:neutrinopredictsplot}
\end{figure}

To explore the correlations in case of $ v_D \sim 10^7\,{\rm GeV}, \Lambda=10^{14}\,{\rm GeV}$, we keep some of the parameters appearing in eqs.~(\ref{eqn:mastereq1}) and (\ref{eqn:mastereq2}) fixed 
while others are varied within specified ranges. 
A comprehensive list of all the parameters and their range is given in tab.~\ref{tab:benchmarkLmuLtau}.
We have chosen the values of the varied parameters ($m_1,\delta_{\rm 
CP},\alpha,\beta, Arg(Y_{\mu,\tau}), Arg(f_{6,7})$) randomly within 
specified ranges and points for which $|f_{3}|,|f_{3^\prime}| \leq 1$ are considered as viable points. We have presented results for normal hierarchy(NH) of neutrino masses. 
A correlation among several parameters can also be seen in 
fig.~\ref{fig:neutrinopredictsplot}. 
The Majorana phase $\alpha$ tends to be either close to 0 or $2\pi$ 
while  $\beta$ shows a peculiar pattern with $\delta_{\rm CP}$. We 
also found that small values of the lightest neutrino mass $m_1$ is 
disfavoured since that will enhance the left-hand side of 
eqs.~(\ref{eqn:mastereq1}) and (\ref{eqn:mastereq2}) which in turn drives $|f_{3}|,|f_{3^\prime}|$ to be larger than unity thus violating perturbativity. 
We found a lower limit of $m_1 \geq 0.0011$eV in our random scan over a billion points. Interestingly, $\delta_{\rm CP}$ is rather unconstrained in this scenario.

Since $m_1$ can not be arbitrarily low, we expect chances of observing neutrinoless double beta decay($0\nu\beta\beta$). 
The amplitude for $0\nu\beta\beta$ decay is given by,
\begin{equation}
\mathcal{A}^{0\nu\beta\beta} \propto \underset{i=e,\mu,\tau}{\sum}U^2_{e i}\;m_i \;\mathcal{M}_{\rm NME}(m_i) + U^2_{\nu\psi}\; M_\psi\; \mathcal{M}_{\rm NME}(M_\psi)
\label{eqn:mbetabeta}
\end{equation}
where $\mathcal{M}_{\rm NME}(\mu)$ is the nuclear matrix element (at the scale $\mu$). We have explicitly checked that the second term in eq.~(\ref{eqn:mbetabeta}) 
(DM contribution) is negligible compared to the first term (active neutrino contribution) due to the smallness of the DM-SM neutrino mixing angle 
$U_{\nu\psi}$(see eq.~(\ref{eqn:DMmix})). In addition, for $M_\psi \gtrsim 100\,{\rm MeV}$ the nuclear matrix element is a sharply decreasing 
function of energy, $\mathcal{M}_{\rm NME}(M_\psi) \ll \mathcal{M}_{\rm NME}(0)$~\cite{Blennow:2010th} . Thus the effective Majorana neutrino mass 
is given by the pure RH neutrino contribution and is related to the light neutrino masses and mixings as,
\begin{equation}
|m_{\beta\beta}| \approx | c^2_{12}c^2_{13} m_1 + s^2_{12}c^2_{13}e^{i\alpha}m_2 + s^2_{13}e^{-i(2\delta_{CP}-\beta)}m_3|.\
\label{eqn:mbetabetaval} 
\end{equation}

\begin{figure}[t]
\begin{center}
\includegraphics[width=9cm]{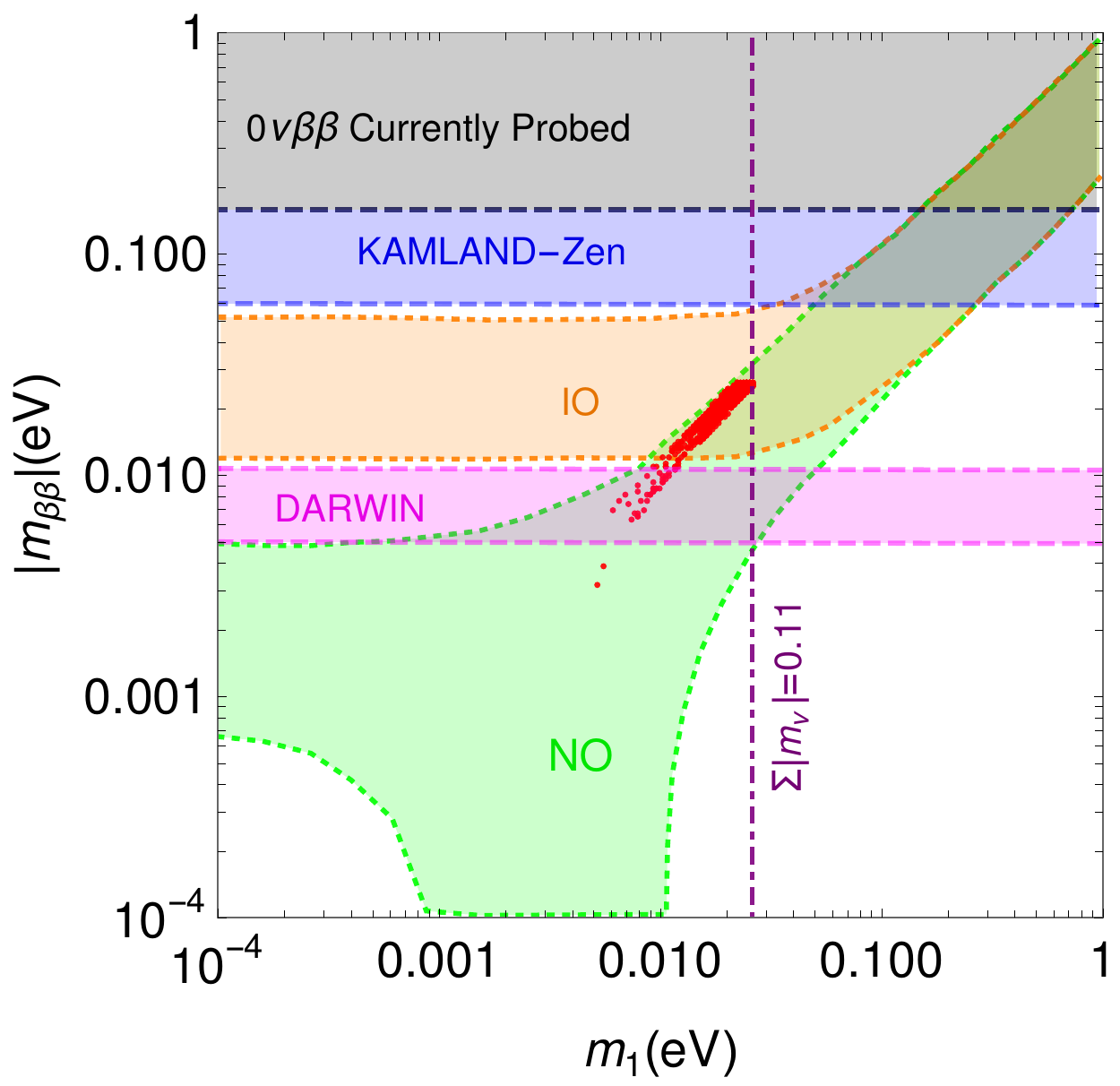}
\end{center}
\caption{ Prediction for $0\nu\beta\beta$ from our model for the parameter points obtained during our scan over a billion points. 
The absence of cancellation between different phases maximizes $|m_{\beta\beta}|$ in our model.} 
\label{fig:0nu2beta}
\end{figure}

We have plotted our prediction for $|m_{\beta\beta}|$ in fig.~\ref{fig:0nu2beta}. We have also showed the projected sensitivity of KamLAND-Zen~\cite{Shirai:2017jyz} and the DARWIN~\cite{Agostini:2020adk} experiements and our model can be probed by DARWIN. 
It is interesting to note that since $m_1 \gtrsim 0.0011\,{\rm eV}$ all the light neutrino masses are nearly of same order in our model. 
Consequently, the third term in eq.~(\ref{eqn:mbetabetaval}) is always negligible compared to the first two terms due to the smallness of 
$s_{13}$ and $m_{\beta\beta}$ is dictated by the sum of the first two terms. 
Moreover, fig.~\ref{fig:neutrinopredictsplot} indicates that most of the solution points are concentrated in the neighbourhood of small $\alpha$ or 
near $\alpha = \,2\pi$, and thus no cancellation among the different contributions can take place giving 
$|m_{\beta\beta}|\simeq c^2_{13}m_1 \approx m_1$, along the upper boundary of the normal hierarchy green band in
fig.~\ref{fig:0nu2beta}.

Since the neutrino masses are essentially generated via Type-I SeeSaw mechanism the mass eigenvalues of the heavy Majorana neutrinos in our model do not depend
very strongly on the exact value of the $U(1)_{L_\mu-L_\tau}$ breaking v.e.v. $v_D$ or the cut-off $\Lambda$. We generally obtain heavy neutrino masses in the range 
$ 10^{-4} - 1000 $ GeV, so those states remain near to the EW scale.


\subsection{Dark Matter Phenomenology} 
\label{sec:DM}
The phenomenology of DM has two important parts namely the production and its decay. 

\subsubsection{Dark Matter Production}

Since the DM has very suppressed couplings, we consider again DM production via the $\textit{freeze-in}$ mechanism. 
But in this model, the $\textit{freeze-in}$ production of DM can take place via the decay $\phi \rightarrow \psi\bar{\psi}$ or via $2\rightarrow 2$ 
scatterings $\phi \phi^\dagger (f\bar{f}) \rightarrow \psi\bar{\psi}$ through non-renormalizable operators~\footnote{We have checked explicitly that 
for most of the parameter region the contribution due to $H\,H\,\rightarrow \psi \bar{\psi}$ is negligible compared to the decay contribution
$\phi \rightarrow \psi\bar{\psi}$. On the other hand as long as $M_\Phi \geq m_H$ the $\textit{freeze-in}$ production occurs above EWSB and 
$h\rightarrow \psi\bar{\psi}$ does not contribute.}. We shall assume that $\Phi$ is in thermal equilibrium during the cosmological evolution which 
is true since it mixes substantially with the SM Higgs. The decay contribution is given by~\cite{Hall:2009bx}:
\begin{equation}
Y_{\rm IR} \approx \frac{136}{4\pi^4} \frac{M_{Pl}}{g^s_* \sqrt{g^\rho_*}}\frac{g_\Phi\; \Gamma_{\Phi \rightarrow \psi \bar{\psi}}}{M^2_\Phi}\int_{x_{min}}^\infty \;dx\; x^3\; K_1(x)
\label{eqn:YIR}
\end{equation} 
where $x_{min}=M_\Phi/T_{RH}$. The scatterings $\phi \phi^\dagger (f\bar{f}) \rightarrow \psi\bar{\psi}$ are mediated by the $L_{\mu}-L_{\tau}$ gauge boson $Z_D$. The contribution to the yield from the scatterings $ij \rightarrow \psi \bar{\psi}$ is given by,

\begin{align}
Y^{ij}_{\rm UV} \approx \frac{g^4_D Q^2_\psi M_{Pl}}{\pi^9 g^s_*\sqrt{g^\rho_*} M_\Phi } \int^\infty_{M_\Phi/T_{RH}}dx \int^\infty_0 dy \frac{y^3 K_1(y)}{(y^2-x^2 M^2_{Z_D}/M^2_\Phi)^2 + x^4 \left(M_{Z_D}\Gamma_{Z_D}/M^2_\Phi \right)^2 } \nonumber\\
\times\,
\begin{cases}
\displaystyle{y \times \frac{30\,Q^2_f}{27} }\hspace{1cm}{\rm when}\,ij\,\,{\rm is}\,\,f\bar{f}
\\
\displaystyle{ (y^2-4x^2)^{3/2} \times \frac{15\sqrt{90}\,Q^2_\Phi}{256}}\\
\hspace{2.5cm}\,{\rm when}\,ij\,\,{\rm is}\,\,\Phi \Phi^\dagger 
\end{cases}\;.
\label{eqn:UVcontribfinal}
\end{align} 
where $\Gamma_{Z_D}=\dfrac{g^3_D v_D}{4\pi}\left( 6 Q^2_f + Q^2_\psi + \dfrac{Q^2_\Phi}{4}\left[1-\dfrac{4M^2_\Phi}{M^2_{Z_D}} \right]^{3/2} \right)$ is the total decay width of the gauge boson $Z_D$. 
\begin{center}
\begin{figure}[t!]
\includegraphics[width=8cm]{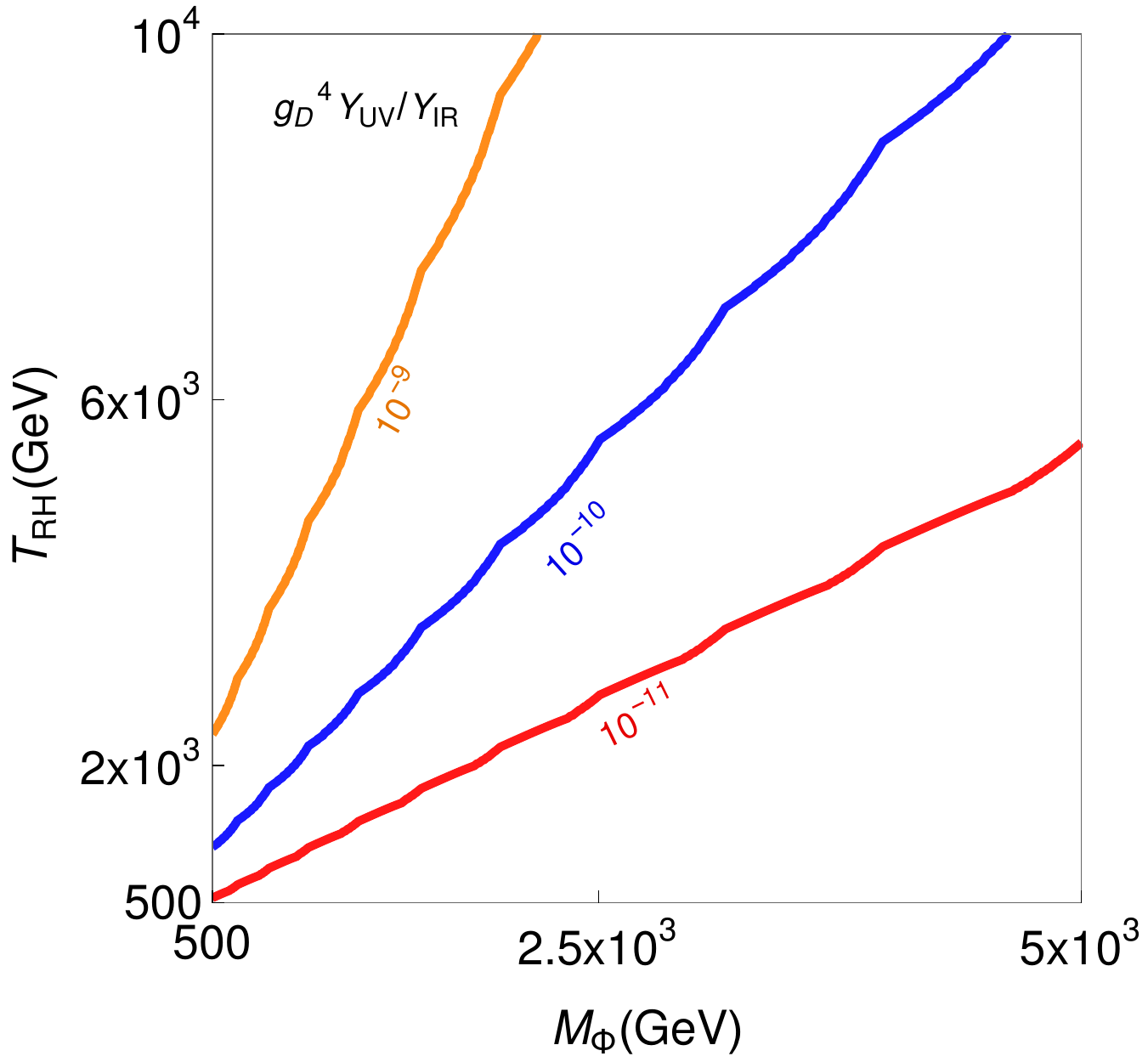}
\includegraphics[width=8cm]{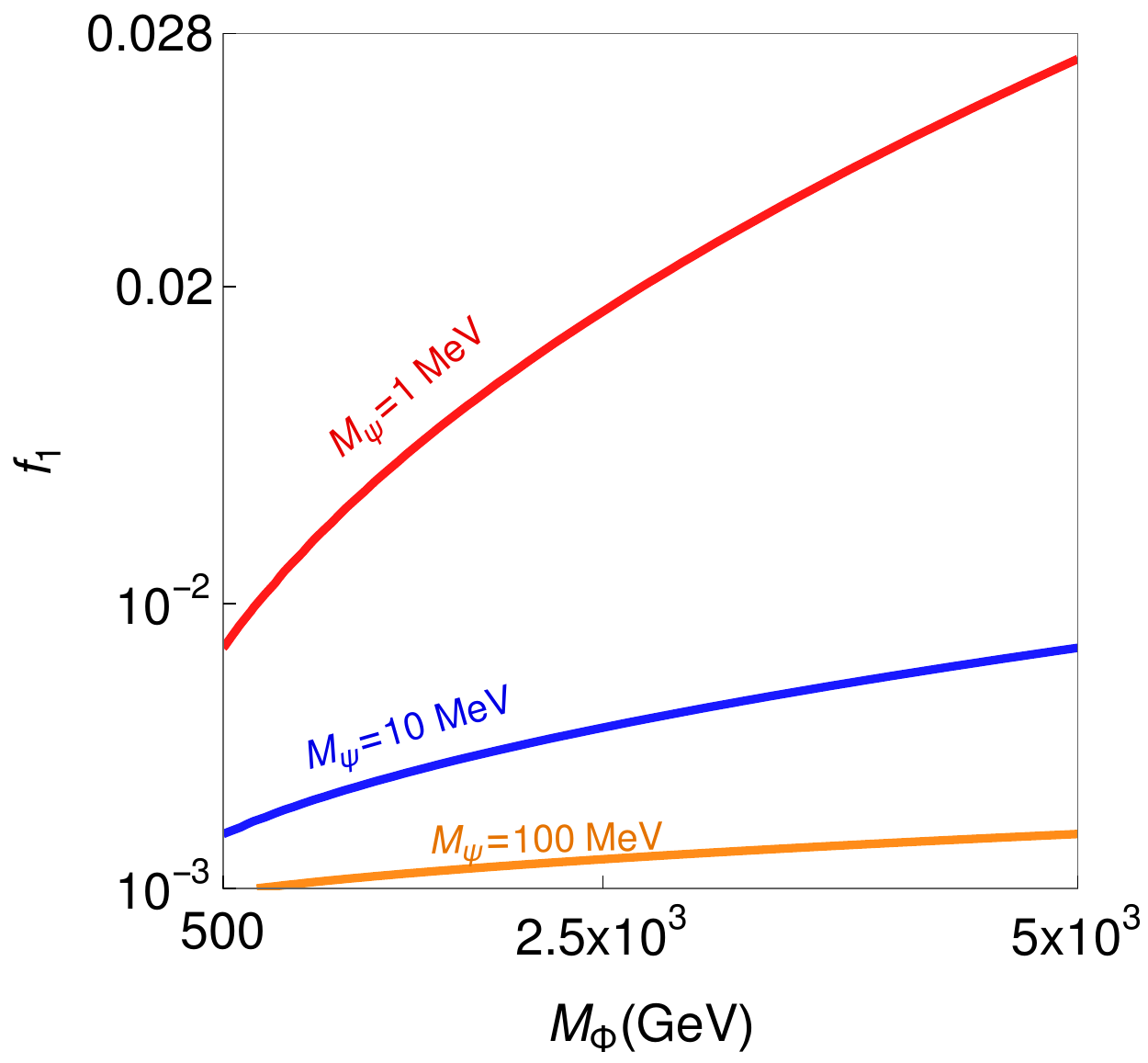}
\caption{Left panel depicts the contours of $g^{4}_DY_{UV}/Y_{IR}$ as a function of $M_\Phi$ and $T_{RH}$. Right panel depicts the $\Omega h^2 = 0.12$ contours in the $M_\Phi-f_1$ plane. }
\label{fig:DMrelicsplot}
\end{figure}
\end{center}
In order to find the relative importance of the decay and scattering process we have plotted the ratio $g^{4}_DY_{UV}/Y_{IR}$ in the {\it left-panel} of 
fig.~\ref{fig:DMrelicsplot}. The contribution from the scattering is negligible as long as $T_{RH}\ll M_{Z_D}$ since the s-channel propagator 
$M_{Z_D}$ is off-shell in this region. Moreover the presence of the $Z_D$ decay width in the denominator gives a $g^{-4}_D$ suppression in 
$Y_{UV}$, which is also evident in fig.~\ref{fig:DMrelicsplot} ({\it left-panel}). 
We have assumed $g_D$ to be 0.01 which ensures that the IR contribution is always dominant.


 Thus considering only the $\phi \rightarrow \psi\bar{\psi}$ contribution we obtain:
\begin{equation}
\Omega_{FI}h^2 \approx 0.1 \left(\frac{f_1 v_D/\Lambda}{1.51\times 10^{-9}}\right)^2 \left(\frac{M_\psi}{1 {\rm MeV}}\right)\left(\frac{M_\Phi}{1 {\rm TeV}}\right)^{-1}\left[1-\frac{4M^2_\psi}{M^2_\Phi} \right]^{3/2}.
\label{eqn:DMfreezeinlmultau}
\end{equation}
The contours of $\Omega_{FI}h^2 = 0.12$ are shown in the right panel of fig.~\ref{fig:DMrelicsplot} in the plane of $f_1$ and $M_\Phi$. 
As the mass of the DM increases, lower 
values of the coupling $f_1$ are required to obtain the correct relic density while $f_1$  can increase for increasing $M_\Phi$. 

\begin{center}
\begin{figure}[t!]
\includegraphics[width=15cm]{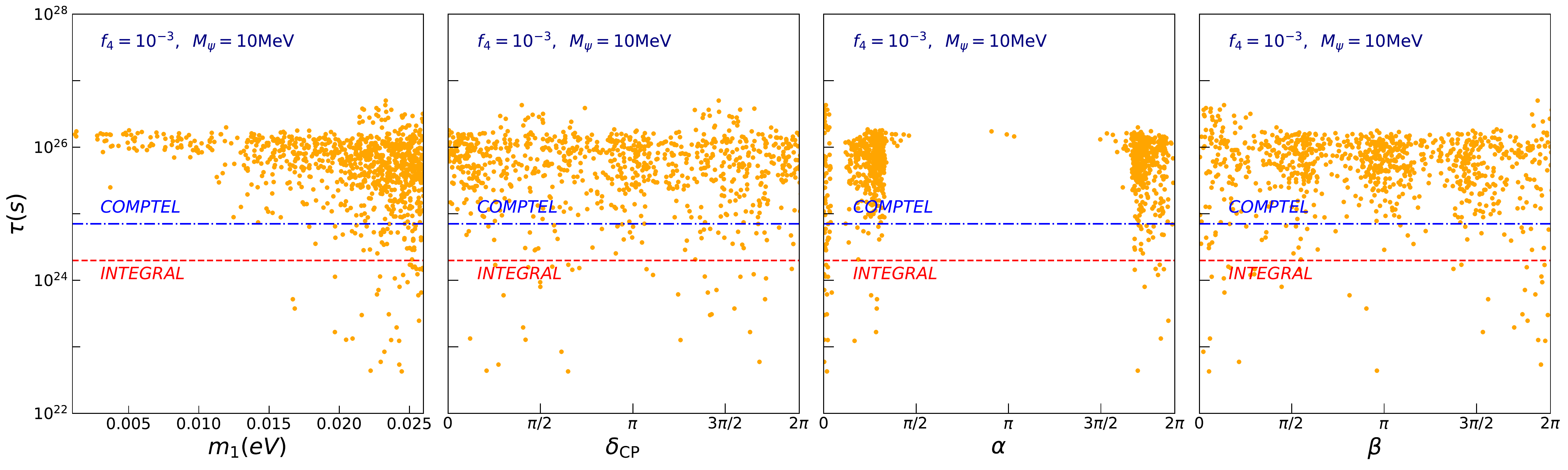}
\includegraphics[width=15cm]{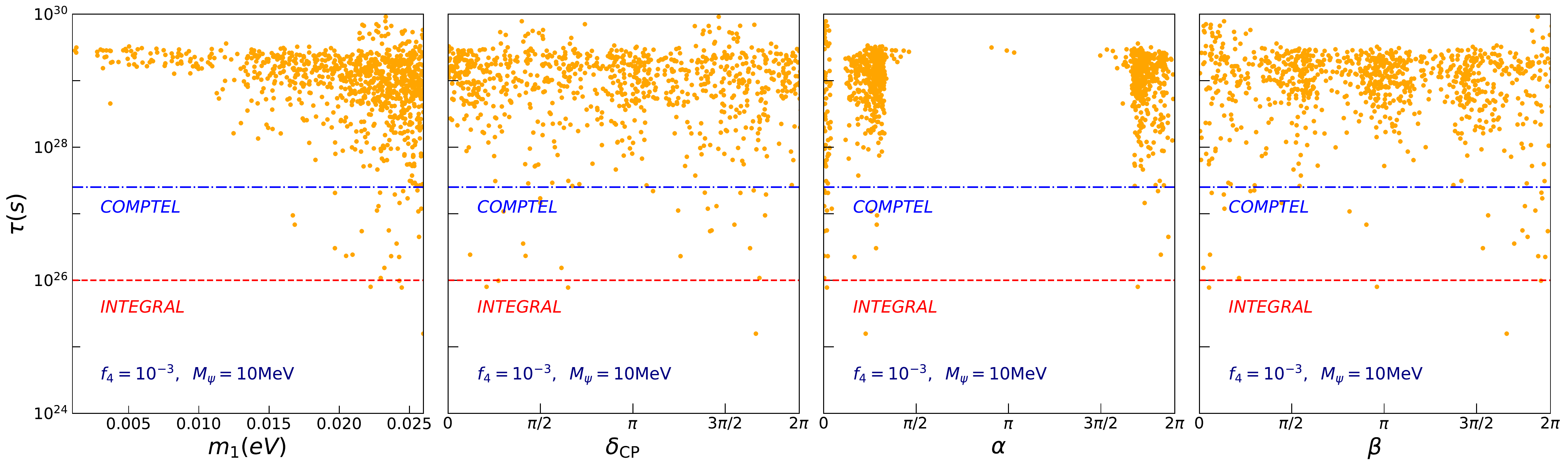}
\caption{Prediction for DM lifetime with several neutrino physics parameters. 
Top panel is for the tree-level decays $\psi \rightarrow l^+l^- \nu, \nu\bar{\nu}\nu$ while bottom panel is for the 
loop-induced decay $\psi \rightarrow \nu\,\gamma$. In both the cases we have shown the available limit from 
INTEGRAL(\textit{dashed dot line}) and COMPTEL(\textit{dashed line}).}
\label{fig:DMdecayplot}
\end{figure}
\end{center}
\subsubsection{Dark Matter Decay}
\begin{center}
\begin{figure}[t!]
\includegraphics[width=7cm]{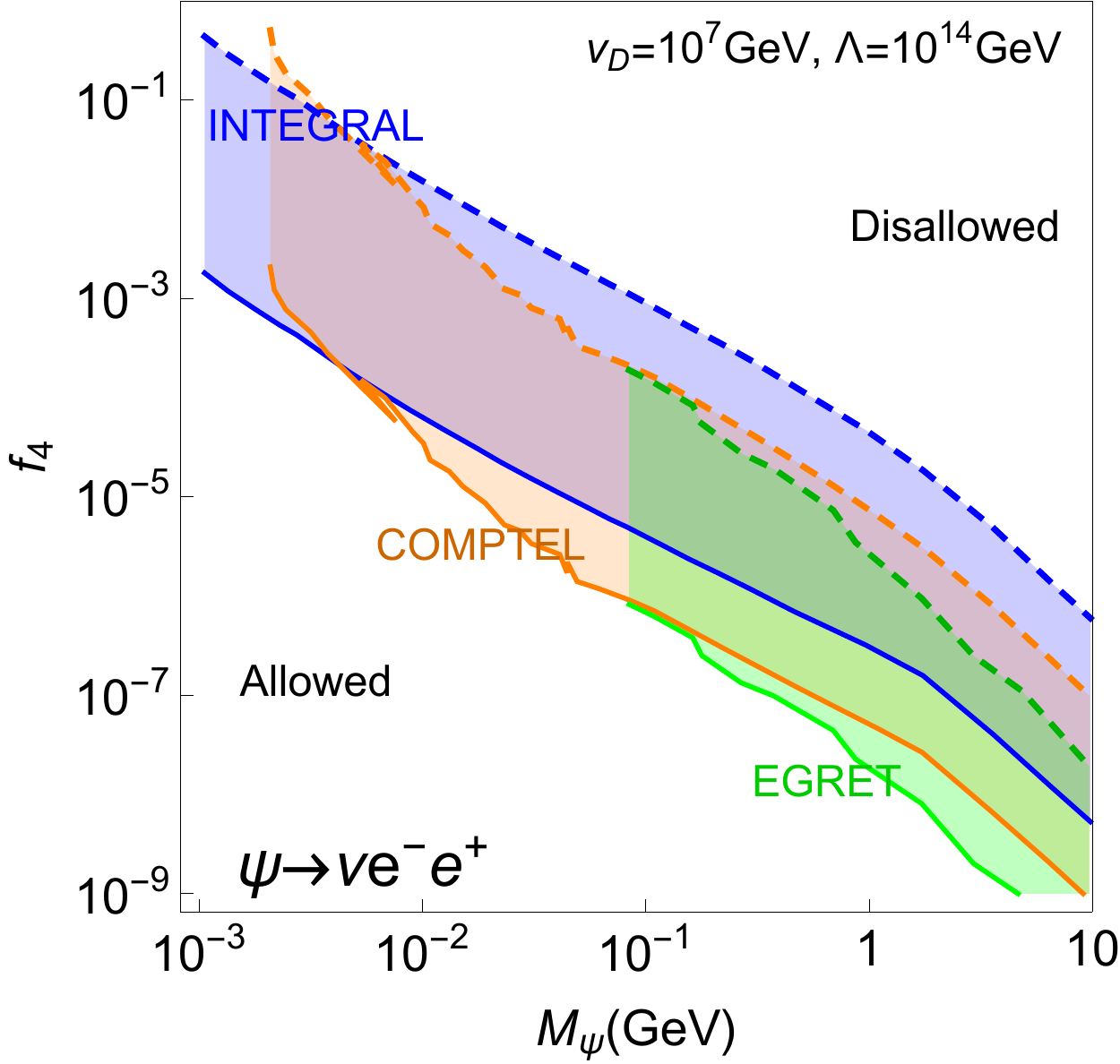}
\includegraphics[width=7cm]{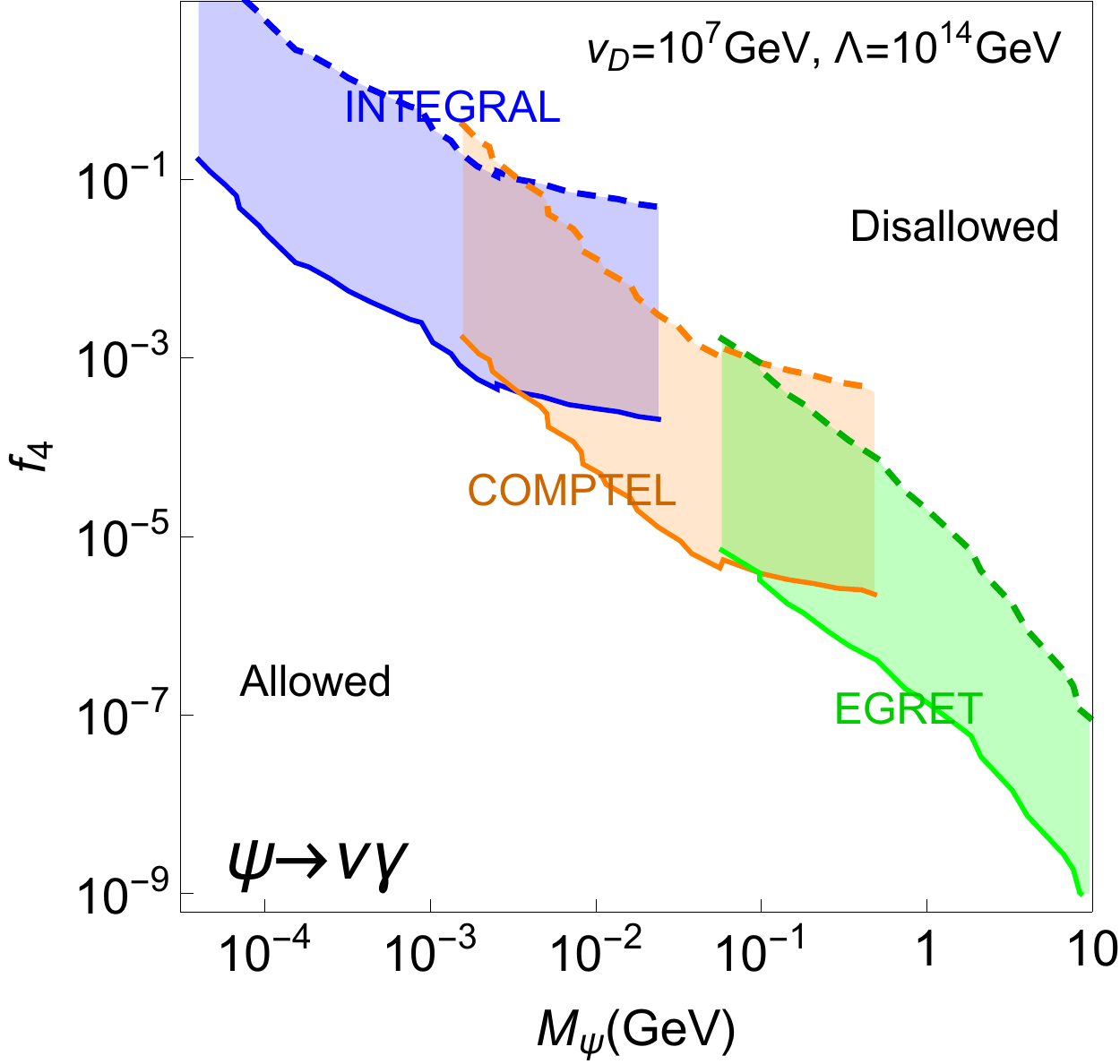}
\caption{Constraints on the $f_4-M_\psi$ plane from INTEGRAL (Blue), COMPTEL (Orange) and EGRET (Green).
The dashed lines correspond to most 
$\textit{conservative limit}$ and the solid lines depict the $\textit{optimistic limits}$. The region above the dashed lines is always ruled out for $v_D=10^7\,{\rm GeV},\Lambda=10^{14}\,{\rm GeV}$.}
\label{fig:DMdecayf4masscons}
\end{figure}
\end{center}

The DM mixes with the SM neutrinos via the mixing given in eq.~(\ref{eqn:DMmix}). 
Thus the DM decays can occur via $\psi \rightarrow Z^* \nu \rightarrow f \bar{f} \nu$ or 
$\psi \rightarrow W^{\pm *} l^{\mp} \rightarrow f \bar{f}^{\prime} l^{\mp}$ process. The corresponding decay width is,
\begin{equation}
 \Gamma_{\psi}^{ff\nu}  \approx  \left(10^{26} s\right)^{-1}\left(\frac{M_\psi}{1 {\rm MeV}} \right)^5 \left(\frac{|U_{\nu\psi}|}{6.345\times 10^{-12}}\right)^2
\end{equation}
and radiative decay of the DM $\psi \rightarrow \nu \gamma$ gives,
\begin{equation}
\Gamma_{\psi}^{\nu\gamma} \approx 
 \left(1.34\times 10^{30} s\right)^{-1}\left(\frac{M_\psi}{1 {\rm MeV}} \right)^5 \left(\frac{|U_{\nu\psi}|}{6.345\times 10^{-12}}\right)^2
\end{equation}
The dependence of DM lifetime on several neutrino physics parameters are shown in fig.~\ref{fig:DMdecayplot} for a DM mass of 10 MeV with 
$f_4 = 10^{-3}$. The horizontal lines are the upper limit on the DM lifetime obtained form INTEGRAL~\cite{Bouchet:2008rp} and 
COMPTEL~\cite{1999ApL&C..39..193W} experiments. As we can see that most of the parameter space is beyond the reach of the present limits. 
In fig.~\ref{fig:DMdecayf4masscons} we have shown the constraints on the  Wilson coefficient $f_4$ vs DM mass $M_\psi$ plane as obtained 
from INTEGRAL~\cite{Bouchet:2008rp}, COMPTEL~\cite{1999ApL&C..39..193W} and EGRET~\cite{Strong:2004de} experiments. The dashed line in 
fig.~\ref{fig:DMdecayf4masscons} depicts the conservative limit (the weakest constraint) on $f_4$, whereas the solid line shows the 
optimistic limit i.e the possible strongest constraint. For our chosen values of $v_D=10^7\,{\rm GeV}, \Lambda=10^{14}\,{\rm GeV}$ 
the region above the dashed line is always disallowed. The benchmark points to derive the conservative and optimistic limit are 
given in tab.~\ref{tab:DMlimittab}.  We found that the constraints on the Wilson coefficient $f_4$ are dominantly determined by 
$\psi \rightarrow e^+e^-\nu$ for $M_\psi > 1\,{\rm MeV}$ in spite of the stronger constraints for the channel in monochromatic 
photons~\cite{Essig:2013goa}. This is mainly because  for the same mixing angle the DM lifetime is nearly four orders of magnitude 
larger in case of the radiative decay due to the loop suppression compared to tree level decay. 
For a DM mass lower than 1 MeV the three body decay is not possible and only radiative decay restricts $f_4$. 
This can be seen from the right panel in fig.~\ref{fig:DMdecayf4masscons} where the DM mass starts from 40 keV.

\begin{table}[t]
\begin{center}
\begin{tabular}{|c|c|c|}
\hline
Conservative  Limit & Parameters & Optimistic Limit\\\hline
 0.023$\,{\rm eV}$ & $m_{1}$ &  $0.026\,{\rm eV}$\\
 1.46$\pi$ & $\delta_{CP}$ & $1.67\pi$\\
 1.99$\pi$ & $\alpha$ &  $0.23\pi$ \\
 1.83$\pi$ & $\beta$ & $1.66\pi$ \\
 $10^{-7}e^{0.049 i\pi}$ & $Y_{\mu}$ & $10^{-7}e^{0.73 i\pi}$ \\
 $10^{-7}e^{1.63 i\pi}$ & $Y_{\tau}$ & $10^{-7}e^{1.36 i\pi}$\\
 $e^{1.5 i\pi}$ & $f_6$ & $e^{0.22 i\pi}$ \\
 $e^{1.1 i\pi}$ & $f_7$ & $e^{1.86 i\pi}$ \\\hline
\end{tabular}
\caption{Neutrino oscillation parameters that gives $\textit{conservative}$ and $\textit{optimistic limits}$ on $f_4$ as shown 
in fig.~\ref{fig:DMdecayf4masscons}.}
\label{tab:DMlimittab}
\end{center}
\end{table}
\subsection{Other Phenomenological Issues}
\label{sec:EDMDD}
\begin{itemize}
\item {\bf{Direct Detection of DM:}} 
The DM $\psi$ carries $ L_\mu - L_\tau $ charge and therefore interacts only with the second and third generation leptons.
Nevertheless, it can scatter off the electrons via $\phi$ and $ h $ mediated t-channel diagrams thus providing the possibility of a signal 
in direct detection at low energies. Following~\cite{Essig:2011nj} one can define a reference cross-section ($\bar{\sigma}_e$) and 
a DM form-factor ($F_{\rm DM}$) as
\begin{eqnarray}
\bar{\sigma}_{e}\,&=&\,\frac{\mu^2_{\chi e}\; y^2_e\; \theta^2_h}{2 m^4_\Phi}\left(\frac{f_1 v_D}{\Lambda}\right)^2\frac{(1+\alpha^2 m^2_e/4m^2_\chi)(1+\alpha^2 m^2_e/4m^2_e)}{\left(1+\alpha^2 m^2_e/m^2_\Phi\right)^2}\\
F^2_{\rm DM}(q)\,&=&\,\frac{(1+q^2/4m^2_\chi)(1+q^2/4m^2_e)}{\left(1+q^2/m^2_\Phi\right)^2}\times \frac{\left(1+\alpha^2 m^2_e/m^2_\Phi\right)^2 }{ (1+\alpha^2 m^2_e/4m^2_\chi)(1+\alpha^2 m^2_e/4m^2_e) }\nonumber\\
\end{eqnarray}
which essentially determine the DM direct detection rate. 
Here $\mu_{\chi e}$ is the reduced mass of DM-electron system and $y_e$ is the electron Yukawa coupling. 
We found that even for $M_\Phi \simeq 10\,{\rm GeV}$ the reference cross-section $\bar{\sigma}_e\,\lesssim\,10^{-69}\,{\,\rm cm}^2$ 
for $\theta_h\,\sim\,0.1$~\footnote{Note that here either in the DM vertex or in the SM vertex a $\Phi-h $ mixing angle must be present. 
and from up-to-date Higgs precision measurement  one has $\theta_h\lesssim 0.1$~\cite{Cheung:2018ave}.}
Such a small value of $\bar{\sigma}_e$ is actually two-fold suppressed: both by (i) the presence of the $\textit{freeze-in}$ coupling $f_1 v_D/\Lambda$ 
and (ii) the electron Yukawa coupling ($y_e$).
A target of muons or tauons would be more promising, as the Yukawa couplings are larger and also the channel with the exchange of a $ Z_D $
gauge boson is possible, but still the expected rate is too low to be measured in future experiments.

\item {\bf Electron Dipole Moment:} The presence of the additional scalar $\phi$ also opens up the possibility of a new
contribution to electron dipole moment(EDM) at two-loop~\cite{Okawa:2019arp}. The EDM contribution is given by, 
\begin{eqnarray}
d^i_{e}= e\,L^i_e\,\int d^4q\,d^4k\,f^i_{\rm scalar}(q,k)
\end{eqnarray}
where we have used $\int d^4q\,d^4k\,f^i_{\rm scalar}(q,k) \simeq 0.95\,M^2_{N,i}/m^2_W$ and 
\begin{equation}
L^i_e\,\simeq\, 4\times 10^{-27}\,{\rm cm}\,\left(\frac{f_{3,3^\prime}v_D}{\Lambda}\right) \theta^2_{\nu N_i} \theta_h \frac{M_{N i}}{m_W},
\end{equation} 
where $\theta_{\nu N_i}$ is the $N_i-\nu$ mixing angle and $f_{3,3^\prime}v_D/\Lambda$ is the vertex factor for $N_i N_i \phi$. 
The smallness of the terms $f_{3,3^\prime}v_D/\Lambda\,\sim\,10^{-7}$ and $\theta_{\nu N_i}\,\sim\,10^{-6}$ gives an enormous suppression 
of order $\sim\,10^{-19}$. As a result we get $d_e\,\lesssim\,10^{-43}\,e\,{\rm cm}$ which is far too low compared to the latest bound from 
ACME ($d_e\,\lesssim\,10^{-29}\,e\,{\rm cm}$)~\cite{Andreev:2018ayy}.  

\item {\bf Collider constraints:}

In this model, the only dark sector fields that may appear at colliders apart from the DM are the
scalar field ($\phi$) responsible for the $ U(1)_{L_\mu-L_\tau} $ breaking and the heavy neutrinos ($N_i$) below the TeV scale. 
Since the scalar mixes with the SM Higgs as in portal models, we expect similar signatures in the 
Higgs sector, i.e. a contribution of DM to the invisible Higgs width and a modification of the Higgs 
couplings to the SM fermions.
Also in case the scalar field is light, it may be produced via the mixing with the Higgs via gluon fusion~\cite{Chang:2017ynj}. For example, $\sigma(gg\rightarrow \phi)\sim 0.5{\,\rm pb}$ for $M_\Phi = 500\,{\rm GeV}$ and $\theta_h=0.1$~\cite{Cepeda:2019klc}.

The heavy Majorana neutrinos couple to the SM via small neutrino Yukawa couplings and to the $\phi$ via Yukawas $h_{e\mu,e\tau}$ etc. 
For our choice of parameters $v_D=10^7\,{\rm GeV},\Lambda=10^{14}\,{\rm GeV}$ we obtain $h_{e\mu,e\tau}\simeq 10^{-7}$ and thus $BR(\phi\rightarrow \bar{N}N) \lesssim 10^{-9}$ even for $\theta_h \simeq 0.1$ and $\bar{N}N$ production rate via $\phi$-mediation is negligible at the 13 TeV LHC. 
On the other hand, the $pp \rightarrow h\rightarrow \bar{N}N$ is also suppressed due to the smallness of $h_{e\mu,e\tau}$ while $pp \rightarrow W^{\pm}\rightarrow \bar{N}l^{\pm}$ is negligible due to smallness of $\nu-N$ mixing $U_{\nu N} \simeq 10^{-6}$. 
At  the LHC with $\sqrt{s}=$13 TeV, even for an integrated luminosity of $\mathcal{L}=3000\,fb^{-1}$, the number of expected $N$ events 
are $ \simeq 0.1$(h-mediation) and $0.01$(W-mediation).
If these heavy sterile neutrinos are produced, they will appear as long-lived states~\cite{Cottin:2018nms,Liu:2019ayx,Chiang:2019ajm}, 
\begin{equation}
c\tau_N \simeq 12\,{\rm km}\left(\frac{M_N}{10\,{\rm GeV}}\right)^{-5}\left(\frac{U_{\nu N}}{10^{-6}} \right)^{-2}.
\end{equation}
On the other hand, in the scenario we discussed, the $Z_D $ gauge boson is very heavy in order to keep the 
dark matter state out of equilibrium, as discussed in section \ref{sec:DM} and therefore does not produce signatures 
at colliders.
\end{itemize}

The overall advantages of this $ U(1)_{L_\mu-L_\tau} $ model are:
\begin{itemize}
\item Non-renormalizable effective operators, suppressed by the cut-off scale $ \Lambda \leq M_{Pl} $  successfully generate 
also in this case the effective couplings involved in DM decay as well as $\textit{freeze-in}$ in the right ballpark, without the
need to fine-tune the Wilson coefficients.  Moreover those operators also contribute to the neutrino masses and allow to
modify the usual $ U(1)_{L_\mu-L_\tau} $ predictions, lowering the sum of the neutrino masses below the present
Planck constraint.

\item For the lowest possible $ v_D $ scale, an interesting cancellation among the different parameters of the neutrino
mass matrix takes place, giving a correlation among the CP phases and the phases of the couplings.
Unfortunately those correlations do not restrict the value of the Dirac phase, but they allow to restrict the range
of the allowed effective mass for neutrinoless double beta decay, pointing to a relatively large value.

\item The scenario is cosmologically consistent and anomaly free.


\item The addition of the $U(1)_{L_\mu-L_\tau} $ breaking scalar $\Phi$ with the mass 
scale it brings in, induces mixings with the Higgs scalar and could allow the production of
the new scalar state at colliders.

\end{itemize}

\section{Summary and Conclusions}
\label{sec:conclusion}
We have studied a set of three models in sequence,  explaining the phenomenology of 
neutrinos and containing a decaying FIMP dark matter candidate. 
We focused  on a fermionic DM as an example. 
To start with, we have considered a simple renormalizable model of
SM singlet fermions which produces from the decay of a vectorlike 
fermion doublet $F$ while decays via Yukawa interactions with SM 
Higgs. The existence of the DM till today requires tiny Yukawa couplings 
of order $\sim 10^{-20}$ or less. Such extremely small couplings albeit 
`$\textit{technically natural}$' are difficult to explain, as well as the presence
of very different Yukawa coupling sizes for neutrinos masses, FIMP
production and DM decay. 

Thus we moved to a model where the DM is charged under an 
additional $U(1)$ under which all SM particles are neutral. In this 
scenario both the DM production and decay occurs via 
higher-dimensional operators and thus are naturally small. These 
models have interesting collider signatures in terms of the 
vectorlike fermion decays, which explain the DM abundance in the 
Universe. Though this model can naturally explain the small 
couplings required for DM production and decay there is no direct
connection between the phenomenology of the neutrino and 
of the DM sectors.
 
Next, we have attributed charges to SM leptons under the added 
$U(1)$. Inspired by the pattern of neutrino mixing as well as  anomaly cancellation, 
the abelian symmetry adopted here is $U(1)_{L_\mu-L_\tau}$. 
We have studied this model in detail and  shown that, due to the
non-renormalizable operators, the neutrino mass matrix is modified.
This makes is possible to satisfy the PLANCK limit 
on the sum of neutrino masses and at the same time  obtain a sizable $0\nu\beta\beta$ rate, 
which could be observed in future generation experiments.
DM production from the decay of the $L_\mu-L_\tau$-charged scalar $\phi$,
playing in this case the role of the mother particle in FIMP production,
has  been computed in detail,  thus eliciting constraints on the 
Wilson coefficients that drive DM decay.

 We have also studied the possibility of DM direct detection via electron 
scattering and the  contribution to electron dipole moment though 
we conclude that these effects are much lower than the reach of the future 
generation  experiments.   Implications  of the neutrino 
physics parameters in DM decay have also been studied. Although
strict correlations are yet to be identified, mostly due to the multiplicity
of parameters, it is expected that further data from the neutrino sector,
especially those on one or more CP-violating phases there, will serve
to validate or restrict a scenario of the kind described here.   

\section{Acknowledgements}\label{sec:Acknowledgements}
The work of AG and BM was partially supported by funding available from the Department of Atomic Energy, Government of India, for the 
Regional Centre for Accelerator-based Particle Physics (RECAPP),  Harish-Chandra Research Institute. 
AG and BM acknowledge the hospitality of Laura Covi and the University of G\"ottingen where important discussions on this project 
took place. LC and TM thank Harish-Chandra Research Institute for visits during the work. 
During the initial stages of this work, LC received funding from the European Union's Horizon 2020 research and innovation
programmes InvisiblesPlus RISE under the Marie Sklodowska-Curie grant agreement No 690575 and 
Elusives ITN under the Marie Sklodowska-Curie grant agreement No 674896.
TM is supported by a KIAS Individual Grant (PG073501) at Korea Institute for Advanced Study.

\providecommand{\href}[2]{#2}
\addcontentsline{toc}{section}{References}
\bibliographystyle{JHEP}
\bibliography{Dmwtz2}

\end{document}